%% file: VSOODT.tex
\documentclass[lettersize,journal]{IEEEtran}
\usepackage{amsmath,amsfonts}
\usepackage{algorithmic}
\usepackage{algorithm}
\usepackage{array}
\usepackage[caption=false,font=normalsize,labelfont=sf,textfont=sf]{subfig}
\usepackage{textcomp}
\usepackage{stfloats}
\usepackage{url}
\usepackage{verbatim}
\usepackage[export]{adjustbox}
\usepackage{graphicx}
\usepackage{cite}
\usepackage{graphicx} 
\graphicspath{{img/}} 
\usepackage{mathtools}
\usepackage{changepage}

\usepackage{textcomp}
\usepackage{microtype}
\usepackage{interval}
\usepackage{booktabs,makecell,tabularx}

\newcolumntype{C}[1]{>{\centering\arraybackslash}p{#1}}
\newcolumntype{L}{>{\raggedright\arraybackslash}X}
\usepackage{siunitx}
\usepackage{etoolbox}
\allowdisplaybreaks
\newrobustcmd{\B}{\bfseries}
\usepackage[dvipsnames]{xcolor}
\usepackage{hyperref}
 
\hypersetup{colorlinks=true,allcolors=[rgb]{1,0,1}}
\usepackage{amsthm}
\theoremstyle{remark}
\newtheorem{remark}{Remark}
\usepackage{stfloats}
\usepackage{textcomp}
\usepackage{balance}
\begin{document}

\title{PI-Controlled Variable Time-Step Power System Simulation Using an Adaptive Order Differential Transformation Method}

\author{Kaiyang Huang,~\IEEEmembership{Student Member,~IEEE, }Yang Liu,~\IEEEmembership{Member,~IEEE, }Kai Sun,~\IEEEmembership{Fellow,~IEEE, }Feng Qiu,~\IEEEmembership{Senior Member,~IEEE. }
\thanks{This work was supported by the Advanced Grid Modeling program of U.S. DOE Office of Electricity under grant DE-OE0000875.}
\thanks{K. Huang and K. Sun are with the Department of Electrical Engineering and Computer
Science, University of Tennessee, Knoxville, TN 37996 USA (e-mail: khuang12@vols.utk.edu; kaisun@utk.edu).}
\thanks{Y. Liu and F. Qiu are with Argonne National Laboratory, Lemont, IL 60439 (e-mail: yang.liu100@anl.gov; fqiu@anl.gov).}
}

\markboth{Published in IEEE Transactions on Power Systems, vol. 39, No. 5, September 2024 (DOI: 10.1109/TPWRS.2024.3361442)}
{Shell \MakeLowercase{\textit{et al.}}: A Sample Article Using IEEEtran.cls for IEEE Journals}


\maketitle

\color{black}
\begin{abstract}
Dynamic simulation plays a crucial role in power
system transient stability analysis, but traditional numerical
integration-based methods are time-consuming due to the small time step sizes.
Other semi-analytical solution methods, such as the Differential Transformation method, often struggle to select proper orders and steps,
leading to slow performance and numerical instability. To address
these challenges, this paper proposes a novel adaptive dynamic simulation approach for power system transient 
 stability analysis. The approach adds feedback control and optimization to selecting the step and order, utilizing the Differential Transformation method and a proportional-integral control strategy to control truncation errors. Order selection is formulated as an optimization problem resulting in a variable-step-optimal-order method that achieves significantly larger time step sizes without violating numerical stability. It is applied to three systems:  the IEEE 9-bus, 3-generator system, IEEE 39-bus, 10-generator system, and a Polish 2383-bus, 327-generator system, promising computational efficiency and numerical robustness for large-scale power system is demonstrated in comprehensive case studies.
\end{abstract}
\color{black}


\section{Introduction}
\IEEEPARstart{M}{odern} Power systems are characterized by their vast scale and intricate composition, including transmission lines, transformers, generators, loads, and controllers. This complex structure, combined with the diverse and dynamic behavior of the system's elements, can result in a wide range of security issues.
\color{black}In contrast to direct stability analysis such as direct methods \cite{Varaiya1985,HsiaoDong1994,chiangdirect,anghel2013algorithmic}, which are based on non-linear system theories such as Lyapunov stability analysis, time-domain simulation-based methods\cite{milano2005open, Khaitan2008, Fabozzi2013, Aristidou2014,stiasny2023solving} are able to provide more detailed and accurate dynamics of a power system for its dynamic security assessment (DSA).\color{black}
 This flexibility allows for a more comprehensive and accurate assessment of system stability, compared to other methods, and enables the identification of potential issues that may be overlooked by other approaches\cite{chiangdirect,anghel2013algorithmic} which are critical for power system operators and engineers in making informed decisions to ensure the system's safe and reliable operation. \color{black} However, while time-domain simulation can provide detailed and accurate dynamics, it is time-consuming due to the large-scale power systems models compared with direct methods. Time-domain simulation involves solving the system's mathematical model described by Ordinary Differential Equations (ODEs) or Differential Algebraic Equations (DAEs) \cite{Milano2010} as an Initial Value Problem (IVP). However, traditional numerical integration schemes used in commercial software, such as Runge-Kutta (RK) methods\cite{dormand1980family} for ODEs, or Runge-Kutta-Newton-Raphson methods \cite{Milano2010} for DAEs often suffer from computational burdens or stability issues \cite{tzounas2022smallsignal, TZOUNAS2022108266,tzounas2023unified}. Therefore, advanced time-domain simulation techniques are required to maintain stability in modern power systems. Therefore, developing advanced time domain simulation techniques is crucial for ensuring the stability and reliability of modern power systems.\color{black}
\par Over the years, significant advancements and improvements have been achieved in time domain simulation to enhance their accuracy, efficiency, and stability. Various advanced simulation techniques have been developed such as simulations based on parallel approaches\cite{ Aristidou2014, Gurrala2016}, high-performance computing\cite{jin2017comparative}, machine learning\cite{stiasny2023solving, Xiao2023} and Semi-Analytical Solution (SAS) methods\cite{wang2018time,liu2019power,liu2019solving,yu2022non}, to provide faster and more accurate simulation results. SAS methods are promising simulation technologies that have recently gained attention. These methods involve dividing the computational burden of solving power system differential equations or algebraic equations into two stages which combine numerical and analytical tools to solve the system's equations, resulting in a more efficient and accurate simulation.
In the precomputing stage, an approximate and analytical solution is derived for the power system differential or algebraic equations model based on different spectrums. This solution is an explicit or implicit expression of state variables, the initial state, and parameters and is accurate in a specific time interval, the length of which depends on the system model and the order of the SAS expression. In the online stage, the SAS is evaluated by solving a precomputing model, which usually transforms IVP to algebraic equations with different orders. This process is repeated until the desired simulation length is reached. A variety of SAS methods have been investigated in the literature, such as the Adomian decomposition method \cite{duan2016power}, the power series method\cite{wang2018time}, the holomorphic embedding method\cite{yao2019efficient}, the differential transformation method\cite{liu2019power,liu2019solving,yu2022non}, etc. These methods have shown promising results in power system simulations and have been tested and validated on large-scale power systems\cite{park2021examination}. Several examinations \cite{liu2019power,park2021examination} show that the Differential Transformation (DT) based time domain simulation has a better speed performance than some RK methods when high accuracy is required for the power system simulation.
\par \color{black}Basically, the order and step size are determined case by case and fixed based on experiences. From \cite{liu2019power}, the DT method is able to increase the step size by using a high-order SAS. And, as a semi-analytical simulation method, the DT method is able to adjust its step size more easily on the fly of the simulation than numerical methods such as RK methods. However, the theoretical linkage between the step size and the order of the SAS, or in other words the optimal pair of step size and the order, has not been studied well. To ensure numerical stability, a step size much smaller than its threshold is often employed in simulations. In fact, how to optimally adjust both the step size and order on the fly of the simulation is not a trivial problem. Variable step and variable order strategies should not be too complex to ensure the efficiency of the simulation, and also the accuracy of the simulation should be acceptable. These requirements will be met by the proposed strategies. \color{black} One advantage of using the DT method is that the recursive computation process is performed offline, resulting in faster online iteration. Moreover, analyzing error estimation is easier when compared to the embedded RK method \cite{dormand1980family}. The DT algorithm can adjust the step size based on truncated error without the need to embed another order's solver for error prediction. This feature is demonstrated in a simple case in \cite{yu2022non}, where an adaptive time window is utilized in a multi-energy system. The aforementioned strategy can be extended to a PI-controlled strategy \cite{PIref}, which performs better in stiff systems and reduces rapid changes in step size. Unlike traditional numerical methods, the DT method can smoothly adjust orders during the simulation, allowing for the design of an adaptive order that works in conjunction with the step strategy to achieve improved performance. 
\color{black}
\par The contributions of this paper are summarized as follows. First, an automatic PI-controlled variable step strategy for the DT method, referred to as the Variable-Step DT (VS-DT) method, is proposed for power system simulations using a fixed order SAS to achieve high-speed performance with good accuracy, even for large systems. Second, an optimally variable order strategy is proposed to enable using the optimal order in the VS-DT method, which is referred to as the Variable-Step-Optimal-Order DT (VSOO-DT) method. By formulating and solving an optimization problem, the optimal SAS order is determined efficiently for the VS-DT method. The variable step and variable order strategies together provide an optimal candidate pair of the step size and order at each time step on the fly of the simulation, ensuring a good balance between numerical stability and simulation speed. Furthermore, the numerical stability of the proposed methods as well as the selections of all parameters are also theoretically analyzed. Finally, by case studies, the two proposed methods are examined through a variety of tests. The test results show, in particular, that compared to similar methods, the proposed VS-DT and VSOO-DT methods achieve more robust and efficient simulations by automatically varying the step size or optimizing the order as well. Moreover, the proposed two strategies suggest a general framework for simulations using adaptive step sizes and optimized solution orders, not limited to the DT method. Such a framework can be easily generalized to other semi-analytical simulation methods or numerical methods whose orders are adjustable.
\color{black}
\par The structure of this paper is as follows. In section II, the DT method is reformulated. In Section III we propose a novel PI-controlled variable step size and variable order DT algorithm and analyze the numerical stability. In Section IV we formulate the changing of orders as an optimization problem and a heuristic method is applied to relax the original problem. Finally, in section V, results of time domain simulation based on the new algorithm for several IEEE classical systems are presented to show the scalability and efficiency of the proposed method. We conclude in Section VI by summarizing several advantages of the new strategy and many promising ways to enhance the approach. \color{black}The detailed power system model is provided in the Appendix.\color{black}


\section{Differential Transformation Method}
Consider a dynamic system which is Lipschitz on $\mathbb{R}^{m_1}$:
\begin{equation}\label{or}
    \dot{\bf{x}}=\bf{f(x),}
\end{equation}
where $\bf{x}$ is in $\mathbb{R}^{m_1}$. Assuming that the solution $\boldsymbol{\phi}{(t,\bf{x_0})}$ is sufficiently smooth in a neighborhood $\mathcal{B}_{\color{black}{\delta}}(t_n,\bf{x_0})$, we can expand it using the Taylor series up to a certain order $K$ within $\mathcal{B}_\delta(t_n,\bf{x_0})$:
\begin{equation}
\label{tay}
\begin{split}
\boldsymbol{\phi}(t,\mathbf{x_0}) &=\sum_{k=0}^{K}{\frac{1}{k!}\frac{\partial^{k}{\boldsymbol{\phi}(t,\mathbf{x_0})}}{\partial{t}^{k}}\bigg|_{t=t_n}(t-t_n)^k}\\
    &+\boldsymbol{R(\zeta)}(t-t_n)^{K+1}, 
\end{split}
\end{equation}
where $\boldsymbol{R(\zeta)}$ is related to order $K$, $t_n$ and $t$. \color{black}Substituting \eqref{tay} into \eqref{or}\color{black}, and leveraging the property of the linear independence of polynomials, we can equate the coefficients of the same order, resulting in:
\begin{equation}\label{rec}
\begin{split}
    (k+1)\mathbf{X}(k+1)&=\frac{\text{d}\mathbf{f}}{\text{d}\mathbf{x}}\mathbf{X}(k)+{\mathbf{F_\textbf{non}}{(k)}},    
\end{split}
\end{equation}
where
\begin{align*}
    \mathbf{X}(k)&={\frac{1}{k!}\frac{\partial^{k}{{\phi}(t,\mathbf{x_0})}}{\partial{t}^{k}}\bigg|_{t=t_n}}
\end{align*}
and $\mathbf{F_\textbf{non}}{(k)}$ is a nonlinear function from $k=0,\cdots,K$. A recursive rule to obtain a semi-analytical solution in a single step is defined in (\ref{rec}). However, when the system dimension becomes large, solving $\mathbf{F_\textbf{non}}{(k)}$ directly can be computationally complicated due to the high-order partial derivatives. Therefore, further modification is required. To lift the original system into a linear form, we define intermediate variables $\mathbf{y}\in\mathbb{R}^{m_2}$ to eliminate $\mathbf{F_\textbf{non}}{(k)}$ in a linear form. Suppose $\mathbf{y}=\mathbf{g}(\mathbf{x})$, then, let $\mathbf{z}=(\mathbf{x}^\top,\mathbf{y}^\top)^\top\in \mathbb{R}^{m_1+m_2}$, we can rewrite $\mathbf{f}$ as an augmented system $\mathbf{f_{aug}}:\mathbb{R}^{m_1+m_2}\to \mathbb{R}^{m_1}$ such that $\mathbf{f_{aug}}\left(\mathbf{x},\mathbf{g}(\mathbf{x})\right)=\mathbf{f}(\mathbf{x})=\mathbf{f_{aug}}(\mathbf{z}),\forall \mathbf{x} \in \mathbb{R}^{m_1}$, by proper selecting intermediate variables, $\mathbf{f_{aug}}(\mathbf{z})$ is an vector linear function with respect to $\mathbf{z}=(\mathbf{x}^\top,\mathbf{y}^\top)^\top$. Specifically, we assume that $\mathbf{y}$ is a vector function of $\mathbf{x}$, and we can derive a DT rule offline for the nonlinear algebraic function $\mathbf{y}=\mathbf{g}(\mathbf{x})$. Using the chain rule of derivatives, we can create more than one layer of variables from $\mathbf{y}$ to $\mathbf{x}$. The iteration rule can be obtained from the Taylor series of \eqref{tay} by both sides due to the linear independence of polynomials:
\begin{equation}\label{new2}
    \mathbf{Y}(k)=\mathbf{G}(\mathbf{X}(k),\mathbf{X}(k-1) ,\cdots ,\mathbf{X}(0)),
\end{equation}
where $\mathbf{G}$ is a DT function from $(\mathbf{X}(k),\mathbf{X}(k-1) ,\cdots ,\mathbf{X}(0))$ to $\mathbf{Y}(k)$ and
\begin{align*}
          \mathbf{Y}(k)&={\frac{1}{k!}\frac{\text{d}^{k}{{\mathbf{y}}(t)}}{\text{d}{t}^{k}}\bigg|_{t=t_n}}.  
\end{align*}
Based above, a closed form of (\ref{rec}) can be obtained:
\begin{equation}\label{rec2}
\begin{split}
    (k+1)\mathbf{X}(k+1)&=\frac{\text{d}\mathbf{f_{aug}}}{\text{d}\mathbf{z}}\mathbf{Z}(k)=\mathbf{J}\mathbf{Z}(k),\\
    \mathbf{Y}(k)&=\mathbf{G}(\mathbf{X}(k),\mathbf{X}(k-1) ,\cdots, \mathbf{X}(0)),\\
    \mathbf{Z}(k) &=(\mathbf{X}(k),\mathbf{Y}(k))
\end{split}
\end{equation}
from $k=0,\cdots,K-1$. The matrix $\mathbf{J}$ remains a constant matrix, while the higher-order nonlinear terms $\mathbf{F_\textbf{non}}{(k)}$ have been eliminated and transformed into a more manageable derived vector function $\mathbf{G}$ with expanded dimensions. In \cite{liu2019power}, various forms of nonlinear functions, such as square functions and sine functions, have been derived. Moreover, the DT iteration formula for the detailed and classical models \eqref{equ:1} for power system dynamic simulations has been proposed in \cite{liu2019power,liu2019solving}. This iteration process takes the form of (\ref{rec2}), providing a $K^{\text{th}}$ order approximation solution of $\boldsymbol{\phi}$ in $\mathcal{B}_\delta(t_n,\bf{x_0})$. Finally, an algorithm can be designed to solve the IVP (\ref{or}):

\begin{algorithm}[H]
\caption{Original DT Method}\label{alg:DT}
\begin{algorithmic}[1]
\REQUIRE ${t_0, }$ ${t_e, }$ ${h, }$ $K,$ ${\mathbf{x_0}}.$
\STATE initialization $t=t_0, $ ${{\mathbf{X(0)}}={\mathbf{x_0}}, }$
\STATE ${{\mathbf{Y(0)}}={\mathbf{G}}({{\mathbf{X(0)}}}}), $$\mathbf{Z}(0) =(\mathbf{X}(0),\mathbf{Y}(0)), $ $n=0.$
\WHILE{$t+h<t_e$}
    \WHILE{$k<K$}
        \STATE  $\mathbf{Y}(k) \gets \mathbf{G}(\mathbf{X}(k),\mathbf{X}(k-1) \cdots \mathbf{X}(0))$
        \STATE $\mathbf{Z}(k) \gets (\mathbf{X}(k),\mathbf{Y}(k))$ 
        \STATE $\mathbf{X}(k+1) \gets {(k+1)^{-1}}\mathbf{J}\mathbf{Z}(k)$
        \STATE $k \gets k+1$
    \ENDWHILE
\STATE $\mathbf{x}(t_{n+1}) \gets \sum^{K}_{k=0}{\mathbf{X}(k)h^k}$
\STATE ${{\mathbf{X(0)}} \gets {\mathbf{x}}(t_{n+1}). }$
\STATE $t_{n+1} \gets t+h, t \gets t+h, n \gets n+1, k \gets 0$
\ENDWHILE
\RETURN $\mathbf{x}(t), $ $t_0\leq{t}\leq{t_e}$
\end{algorithmic}
\end{algorithm}
\color{black}
\begin{remark}
Algorithm \ref{alg:DT} provides a general way to implement the DT method into dynamic simulations on an ODE model. For a more general DAE model, a similar process can be implemented \cite{liu2019solving}. In such cases, different from iterative numerical methods, a DT-based dynamic simulation is conducted in a recursive manner using a high-order power series, i.e., the SAS, in which higher-order mathematical terms can be recursively derived from low-order terms. Thus, a high-order approximation of the true solution is achieved, compared with traditional low-order numerical methods, e.g., the RK4 method. Furthermore, there are two main approaches to solving the DAE model: simultaneous and partitioned methods. Simultaneous methods handle both differential and algebraic equations together, while partitioned methods alternate their solutions. While the partitioned method is widely used, solutions based on partitioned methods can face numerical issues due to the interfacing of different variables and it is difficult to use large time step sizes. In contrast, benefiting from the non-iterative and recursive nature of the SAS, the numerical error induced by the partitioning \cite{tzounas2023unified} can be avoided by the DT method.
\end{remark}
\begin{remark}
While higher order accuracy and numerical stability can be provided by the DT method on a power system modeled by DAEs or ODEs, the resulting solution of \eqref{or} is supposed to be continuously differentiable in a desired order for the existence and convergence of the power-series approximation. In general, this assumption can easily be met considering the smooth response of state variables in a power system model even with switchable controllers considering control criteria or limits. The vector field $\bf{f(x)}$ of the power system model is at least piecewise smooth, so the DT algorithm is still well-defined and the corresponding SAS can be derived for each smooth sub-region of the vector field in the power series form of $\mathbf{x}(t_{n+1}) = \sum^{K}_{k=0}{\mathbf{X}(k)h^k}$. Then, the SAS is evaluated over consecutive time intervals of $h$ until making up the desired simulation period. When a switch is detected with any controller during the simulation, the timing can be identified to decrease the step size around the switching point if the resulting error exceeds a threshold based on defect control \cite{ENRIGHT2000159}. Therefore, switches with controllers do not have a significant influence on the time performance of the DT method.
\end{remark}
\color{black}

\section{PI Control-Based Step Strategy}
\subsection{A typical step change algorithm }
In power system simulation, achieving high speed is crucial as online screening requires faster-than-real-time simulation to provide fault information in time, enabling strategies to be implemented promptly. Additionally, simulating a large power system, both in the time domain and for cascading failures, can be exceedingly time-consuming, with no promising solutions yet for dealing with such high-dimensional problems. Therefore, balancing solver accuracy and speed is always a key issue in power system research. 
Adopting a proper step control algorithm for the DT method could be significant for speeding up simulations. One famous step strategy for RK methods addresses step issues by estimating the error in every step and controlling the step size with a given tolerance value. 
\color{black}When a $K^\text{th}$ order solution with the DT method is used, the absolute local truncation error of ${\bf{[x]}}_i$ can be estimated using the last term ${\bf{[X]}}_i(K)h^{K}$. Note that for a power system model, state variables have different physical meanings and units. Therefore, errors are compared across different variables, and they need to be normalized. Unlike the existing work \cite{dormand1980family, yu2022non}, the error index with order $K$ is estimated as follows. Suppose that the approximate solution converges and satisfies the root criteria \cite{lang2003complex}, i.e., there exist a radius $\Bar{r}$ and $\Bar{K}$ determined by $h$ such that $\left(\lVert \mathbf{X}(K)\rVert_\infty h^K\right)^{1/K}\leq\Bar{r}<1\mathrm{~for~}\forall K\geq \Bar{K}$. Equivalently, if $\lVert \mathbf{x}\rVert_\infty$ is bounded, there exist radius $r$ and $\hat{K}$ such that 
\begin{equation}
\label{error2}
r(k):=\left(\max \limits_{1\leq i \leq m_1}\left|\frac{{\bf{[X]}}_i(K)h^{K}}{|{\bf{[x]}}_i|+\eta_i}\right|\right)^{\!{1/K}} \leq r
\end{equation}
for $\forall K\geq \hat{K}$. Here, $\eta_i$ is a small constant introduced to scale the $i^{\text{th}}$ component of $\mathbf{x}$. This is to avoid the singularity caused by $\mathbf{[x]}_i=0$. It also helps adjust the trade-off between absolute and normalized errors. Note that the left-hand side of \eqref{error2} represents the proposed comprehensive error index at order $K$. Once this error at a particular order is determined, the radius $r$ can be approximated using the left-hand side of \eqref{error2}. Then, truncation errors related to higher orders can be bounded using geometric sequences $\sum_{l=K+1}^{\infty} r^{l}$ based on the value of $r$, as per our assumption. Consequently, the local truncation error can be estimated as follows:
\begin{align}
\begin{split}
\label{series error}
    \mathcal{E}\left(K,h\right) &:= \sum_{l=K+1}^{\infty}\left(\max \limits_{1\leq i \leq m_1}\left|\frac{{\bf{[X]}}_i(l)h^{l}}{|{\bf{[x]}}_i|+\eta_i}\right|\right) \\	&\approx\sum_{l=K+1}^{\infty} r^{l}=\frac{r^{K+1}}{1-r}.
\end{split}
\end{align}
In comparison to truncation error estimation based on a single term, \eqref{series error} gives a more conservative error estimation considering higher order terms. \color{black} By setting proper tolerance $Tol$ for per step size error, a control strategy \cite{na1} can be
\begin{align}
\label{s1}
\begin{split}
    e_n&=\frac{\mathcal{E}\left(K,h_n\right)}{h_n},\\
    \theta_{n+1}&=\min\left\{\gamma\left(\frac{Tol}{e_n}\right)^{K^{-1}},\theta_{\text{max}}\right\},\\
    h_{n+1}&={\theta_{n+1}h_n},
\end{split}
\end{align}
where $n$ represents the step stage, from now we simplify $t_n$ as subscript $n$, \color{black}$\theta$ is a dynamic parameter adjusted at each step to determine the next step size, balancing the need for accuracy (keeping $e_n$
  small) and efficiency (allowing for larger step sizes $h_n$ when possible). \color{black} The scheme introduced above provides a simple method for determining a step size that is as large as possible without exceeding the error tolerance, i.e., ensuring that $e_n<Tol$. To reduce the likelihood of a rapid increase in error, a safety factor $\gamma<1$ can be used. This approach has been shown to be quite effective for general problems and can even expand the stability region \cite{na1}. However, it is prone to step oscillations when the system becomes stiff, making the scheme unstable. We will show later that the algorithm above can be considered an integral controller, and that a PI controller typically outperforms it. 
\subsection{A PI Control Based Step Strategy}
In this subsection, a DT scheme based on PI control is applied to accelerate time domain simulations. For simplicity, ignore the practical setting of (\ref{s1}), i.e. the maximum gain constraint $\theta_{\text{max}}$, and take its logarithm to obtain:
\begin{equation}
\label{s2}
\log{h_{n+1}}-\log{h_{n}}=\frac{1}{K}\left(\log{\gamma^KTol}-\log{e_{n}}\right),
\end{equation}
which represents a discrete integral loop. The left-hand side is the difference form of the output derivative with $t$ replaced by $n$, while the right-hand side is the difference between the input $\log{\gamma^KTol}$ and the feedback variable $\log{e_{n}}$. From a control perspective, a PI controller can perform better than a single I controller. Therefore, we consider a more general and flexible framework in this paper, and the stability and robustness of this approach have been proven in \cite{PIref, Hairer2002SolvingOD}, adding a  proportional loop to \eqref{s2}, yields:
\begin{equation}
    \label{s3}
    \begin{split}
        \log{h_{n+1}}-\log{h_{n}}&={K_I}d_n
        +{K_P}\left( d_n-d_{n-1} \right),
    \end{split}
\end{equation}
where $d_n=\left(\log{\gamma^KTol}-\log{e_{n}}\right)$ and $K_I$, $K_P$ are parameters for the integral loop and proportional loop. Rearranging \eqref{s3}, the explicit control strategy can be
\begin{equation}
\label{PIctl}
h_{n+1}=h_n\left( \gamma^K\frac{Tol}{e_{n}}\right)^{K_{I}} \left( \frac{e_{n-1}}{e_{n}}\right)^{K_{P}},    
\end{equation}
It turns out that the typical step change algorithm is a specific case of the PI algorithm with $K_{I}+K_{P}=1/K$ and $K_{P}=0$.
\color{black}

\subsection{Stability of the PI Controller and Selection of Parameters}
The stability of the PI controller is analyzed in this section. Consider the Dahlquist test equation $\dot{x}=\lambda x,~\lambda\in \mathbb{C}$, a standard system for analyzing numerical stability. By adopting the DT method and PI-controlled variable step strategy, the corresponding discrete equations become:
\begin{align}
    \label{dissyt1}
x_{n+1}&=D(z)x_{n}\\
\label{dissyt2}
e_{n}  &=|E(z)x_{n}|\\
\label{dissyt3}
h_{n+1}&=h_n\left( \frac{Tol}{e_{n}}\right)^{K_{I}} \left( \frac{e_{n-1}}{e_{n}}\right)^{K_{P}}
\end{align}
where $z$$:=\lambda h_n$, $D\left(z\right)$$=\sum_{p=0}^{K}{z^p/p!}$, and $E\left(z\right)$$={z^{K}/K!}$. Notice that different from the definition of $e_n$ in (\ref{s1}), the error is analyzed in a general form, i.e., the absolute error, which does not influence the stability analysis.  Difference equations (\ref{dissyt1}) and (\ref{dissyt2}) discrerize $\dot{x}=\lambda x$ based on the DT method considering $K^{\text{th}}$ order accuracy. In (\ref{dissyt2}), the local truncation error is estimated by the last term in the power series, and (\ref{dissyt3}) is rearranged from (\ref{s3}). For the purpose of convergence analysis, let $\alpha=K_{I}+K_{P}$, $\beta=K_{P}$, $\varphi_n=\log{|x_{n}|}$, and $\chi_n=\log{h_n}$, the system above (\ref{dissyt1})-(\ref{dissyt3}) becomes:
\begin{align*}
\varphi_{n+1}&=\log{|D(\lambda e^{\chi_{n}})|}+\varphi_n\\
\chi_{n+1}&=\chi_n-\alpha\log{|E(\lambda e^{\chi_{n}})|}-\alpha\varphi_n\\
&+\beta\log{|E(\lambda e^{\chi_{n-1}})|}+\beta\varphi_{n-1}+\gamma_1
\end{align*}
where $\gamma_1$ denotes some constant, the stability of the system can be analyzed from the radius of the Jacobian matrix, i.e., the transformed matrix of the map: $\left(\varphi_{n},\chi_{n},\varphi_{n-1},\chi_{n-1}\right)\xrightarrow{\text{DT with PI}
}$$\left(\varphi_{n+1},\chi_{n+1},\varphi_{n},\chi_{n}\right)$:
\begin{align}
\label{J}
\frac{\partial{\left(\varphi_{n+1},\chi_{n+1},\varphi_{n},\chi_{n}\right)}}{\partial{\left(\varphi_{n},\chi_{n},\varphi_{n-1},\chi_{n-1}\right)}}=\left(
\begin{array}{cccc}
1 & u & 0 & 0\\
-\alpha & 1-\alpha v & \beta & \beta v\\
1 & 0 & 0 & 0\\
0 & 1& 0 & 0
\end{array} \right)
\end{align}
where $u$ and $v$ are defined by
\begin{align*}
u =\Re\left(\frac{dD(z)}{dz}\cdot\frac{z}{D(z)}\right)\ , v =\Re\left(\frac{dE(z)}{dz}\cdot\frac{z}{E(z)}\right)
\end{align*}
where $\Re(\cdot)$ means the real part of a variable. The Jacobian matrix can be evaluated at the equilibrium point of the mapping, i.e., at $\left(\varphi_{n},\chi_{n}\right)$ such that $|D(\lambda e^{\chi_{n}})|=1$, and $\varphi_{n}$ can be determined from the second equation. By choosing proper values for $\alpha$ and $\beta$, the PI control algorithm (\ref{dissyt3}) expands the stability boundary and increases the simulation speed. Some cases for RK methods are demonstrated in \cite{Hairer2002SolvingOD}. It is known that the DT method shares the same stability domain with the RK method. Thus, the PI control strategy can also expand the stability region of the DT method. However, how to select the proper parameters $K_{P}$ and $K_{I}$ of the PI controller, is an interesting problem. Different from the RK method, the DT method can easily estimate its error by plugging $E\left(z\right)$$={z^{N}/N!}$ into (\ref{s3}) to obtain the linear difference equation if $h_n$ is small:
\begin{equation}
    \begin{split}
        \log{h_{n+1}}-(K\alpha-1)\log{h_{n}}-N\beta\log{h_{n-1}}&=\gamma_2\\
    \end{split}
\end{equation}
where 
\begin{equation}
        \gamma_2=(\alpha-\beta)\log\left(\frac{K!Tol}{\lambda^K} \right)
\end{equation}
The corresponding characteristic equation is
\begin{equation}
\label{ca}
    \begin{split}
        \lambda^2-(K\alpha-1)\lambda-K\beta=0.
    \end{split}
\end{equation}
It turns out that a typical variable step algorithm is a special case of the proposed PI control algorithm when $\alpha=1/K$ and $\beta=0$. If $e_n$ in (\ref{s1}) for each step is estimated using the last term, a slight modification to (\ref{ca}) is required 
\begin{equation}
\label{ca2}
    \begin{split}
        \lambda^2-(K\alpha-1-\alpha)\lambda-(K-1)\beta=0.
    \end{split}
\end{equation}
The selection of $\alpha$ and $\beta$, or in other words $K_I$ and $K_P$ of the PI controller, should make that the modulus of solutions of (\ref{ca}) or (\ref{ca2}) $< 1$ to accelerate the simulation. Also, the radius of the Jacobian matrix in (\ref{J}) needs to be $< 1$ to expand the stability domain. The setting in\cite{controltheory} is followed, which is verified by both control theory and computational case studies. Thus, let $K_I =0.3/K$ and $K_P =0.4/K$. $K_I$ and $K_P$ can be calculated once the order of the DT method is determined.

\color{black}
\section{Optimal Order Selection}
\subsection{The Complexity of Power System Simulation Using DT}
Previous studies \cite{optimalorder, VSVO} focused on selecting the optimal order for the Taylor series, while in this work, we propose a dynamic order selection approach based on the PI and typical step change algorithms. The complexity of the DT method is quantified as the number of multiplication operations performed in each time step. For a classical power system model without controllers, the computational complexity is expressed as:
\begin{align}
\label{complex}
        C(K)=C_\text{ca}(K)=N_{\text{ca}1}K^2+N_{\text{ca}2}K+N_{\text{ca}3},
\end{align}
where $ N_{\text{ca}1}=2N_g$, 
        $N_{\text{ca}2}=2N_g+4{N_g}^2+N_xN_z+N_x$,
        and $N_{\text{ca}3}=N_x$, $N_g$ denotes the number of generators, $N_x=m_1$ is the dimension of the original system (\ref{or}), and $N_z=m_1+m_2$ is the dimension of the augment variable $\mathbf{z}$. The computation of complexity can be divided into three steps. First, we count the multiplication operations involved in calculating the augment variables $Y(k)$ up to $K$ in (\ref{new2}). Second, we calculate the multiplication operations required for computing $X(k)$ up to order $K$. Last, we count the multiplication operations needed for calculating from $X(k)$ to $\mathbf{x}_{n+1}$. 
\par The same procedure can apply to other power system models. For instance, for a detailed model given by (\ref{equ:1a}) to (\ref{detailed}), a complexity function is given by:
\begin{equation}
            \label{complex_detailed}
    C(K)=C_\text{de}(K)=N_{\text{de}1}K^2+N_{\text{de}2}K+N_{\text{de}3},
\end{equation}
 where $ N_{\text{de}1}=17N_g/2$, $N_{\text{de}2}=4{N_g}^2+27N_g/2+N_xN_z+N_x$, and $N_{\text{de}3}=N_x$. With a formula like (\ref{complex}) or (\ref{complex_detailed}) in mind, we can design a special algorithm for optimal order switching based on different power system models. The process of computing complexity is based on the steps outlined in lines 5 to 10 of Algorithm~\ref{alg:DT}.
\subsection{Algorithm of the Adaptive Order Strategy}
From now we use $K_{n+1}$ to represent the order of the DT method used in the step $n+1$. To select the optimal order, we formulate an optimization problem based on the complexity of the simulation model. Specifically, the goal is to maximize the step size $h_{n+1}$ while minimizing the complexity $C(K_{n+1})$, without significantly affecting the accuracy of the simulation. We  model the problem as follows:
\begin{align}
\label{P}
\max \quad \frac{h_{n+1}}{C(K_{n+1})}\
\text{s.t.}\quad (\ref{s1}) \ \text{or}\ (\ref{PIctl}).
\end{align}
Here, note that to strike a balance between stability and speed, as we discussed in the previous section, the PI controller parameters, namely $K_I$ and $K_P$, are dependent on the order of the DT algorithm, and the error estimation is also linked to the choice of order. Given the fact that this optimization problem cannot be solved explicitly, we propose a heuristic approach to separate this problem into three situations: 
\par\textbf{Situation 1.} If the step size has increased or the order has decreased in the previous step, which basically indicates that the error is lower compared to the tolerance, then the order is likely to decrease further in this step to accelerate the simulation. The corresponding condition is:
\begin{align}\label{decrease_condi}
        h_{n-1}\leq h_{n}\text{ or } K_{n}\leq K_{n-1},
\end{align}
if the (\ref{decrease_condi}) holds, the problem is transformed into checking whether the inequality
\begin{align}
\label{decrease}
\frac{h_{de}}{C(K_{de})}>\mu_{de}\frac{h_{es}}{C(K_{n})}
\end{align}
holds, where
\begin{align}
       \label{decrease_al} &K_{de}:=K_{n}-\delta K,\\
       &r_{de}:=r(K_{de},h_n)=\left(\max \limits_{1\leq i\leq m_1}\left|\frac{{\bf{[X]}}_i(K_{de})h_n^{K_{de}}}{|{\bf{[x]}}_i|+\eta_i}\right|\right)^{1/K_{de}},\\
        &e_{de}:=\frac{\mathcal{E}\left(K_{de},h_n\right)}{h_n}=\frac{r_{de}^{K_{de}+1}}{(1-r_{de})h_n},\\ 
        \label{decrease_limit1}&\theta_{de}:=\min\left\{\gamma\left(\frac{Tol}{e_{de}}\right)^{1/K_{de}},\theta_{\text{max}}\right\},\\
        \label{decrease_limit2}&h_{de}:=h_{n+1,K_{de}}=\max\left\{\theta_{de} h_{n},h_{\text{min}}\right\}.
\end{align}
$\mu_{de}$ controls the switching threshold, $\delta K$ is the order intended to decrease in the next step, and $h_{de}$ is the step size candidate obtained from (\ref{s1}) with order $K_n-\delta K$. If $\mu_{de}>1$ and the operation point $\left(h_{de},K_{de}\right)$ performs better than $\left(h_{n},K_{n}\right)$, the order and step are switched which means a reduced order can be used if the same accuracy can be guaranteed to speed up the simulation. The step size $h_{es}$ is estimated by (\ref{s1}) and controlled by a PI controller (\ref{PIctl}). If (\ref{decrease}) does not hold, $h_{n+1}$ is estimated using the PI controller and $K_{n+1} = K_n$ is fixed. $\theta_{\text{max}}$ and $h_{\text{min}}$ are the limits for $\theta_{\text{de}}$ and $h_{\text{de}}$, respectively.

\par\textbf{Situation 2.} If the step size has decreased or the order has increased in the previous step, which indicates that the error is larger compared with tolerance, then the order tends to increase in this step to mitigate the error. The condition is 
\begin{align}\label{increase_condi}
        h_{n}\leq h_{n-1}\text{ or } K_{n-1}\leq K_{n}.
\end{align}
Its satisfaction transforms the problem to checking whether this inequality holds:
\begin{align}\label{increase}
        \frac{h_{in}}{C(K_{in})}>\mu_{in}\frac{h_{es}}{C(K_{n})}
\end{align}
where
\begin{align}\label{increase_al}
&K_{in}:=K_{n}+\Delta K,\\
\begin{split}
\label{increase_rho}  &\rho_{in}:=\min \\ &\left\{ \max \limits_{1\leq i\leq m_1}\left|\frac{{\bf{[X]}}_i(K_n-1)}{{\bf{[X]}}_i(K_n)}\right|,
\max \limits_{\substack{1\leq i\leq m_1\\0\leq j\leq 1}}\left|\frac{{\bf{[X]}}_i(K_n-(j+2))}{{\bf{[X]}}_i(K_n-j)}\right|^{\frac{1}{2}}\right \} ,
\end{split}
\\
\label{increase_est}&e_{in}:=\frac{e_{n}}{\rho_{in}}=\frac{\mathcal{E}\left(K_{n},h_n\right)}{\left(\rho_{in}{h_n}\right)^{\Delta K}},\\
&\theta_{in}:=\min\left\{\gamma\left(\frac{Tol}{e_{in}}\right)^{K_{in}^{-1}},\theta_{\text{max}}\right\},\\
\label{increase_last}&h_{in}:=h_{n+1,K_{in}}=\min\left \{ \theta_{in} h_{n},h_{\text{max}}\right \},
\end{align}
$\mu_{in}$ is the threshold for switching orders. Different from \textbf{situation 1}, the higher order truncation error is implicit as the coefficients' vector ${\bf{X}}(K_n+\Delta K)$ is unknown for a $K_n^{\text{th}}$ order DT method. So a convergence rate factor $\rho_{in}$ approximates the local truncation error candidate $e_{in}$ that corresponds to the order $K_{in}$, and \color{black}the ratios in \eqref{increase_rho} involving different orders help approximate how much the solution has changed between orders. By considering different ratios, potential issues caused by even or odd functions can be avoided. Furthermore, taking the minimum of these ratios ensures that the convergence rate factor is conservatively estimated, providing a safety margin for the increase in order. \color{black}$\Delta K$ represents the increased order in the $n^{\text{th}}$ step, and $h_{in}$ represents the candidate step size predicted by (\ref{s1}) with order $K_{in}$. If (\ref{increase}) holds, the operation point switches from $\left(h_{n},K_{n}\right)$ to $\left(h_{in},K_{in}\right)$ since a higher order DT method enables a larger time step at step $n+1$. Set $K_{n+1}=K_{n} $ and $h_{n+1} = h_{es}$ by (\ref{PIctl}) if (\ref{increase}) doesn't hold.

\textbf{Situation 3:} If conditions for \textbf{situation 1} and \textbf{situation 2} are both satisfied, then the order is determined by optimizing the objective function with respect to all three values. For simplicity, assume that $\mu_{in}=\mu_{de}=1$. \color{black}If the conditions $(\ref{decrease_condi})\text{ and }(\ref{increase_condi})$ hold, the problem is equivalent to solving\color{black}
\begin{align}\label{situation3}
\begin{split}
\left(h_{n+1},K_{n+1}\right)\in&\arg\max\limits_{\substack{\left(h,K\right)\in\mathcal{I}}}\frac{h}{C(K)},
\end{split}
\end{align}
such that $\mathcal{I}=\left\{\left(h_{de},K_{de}\right),\left(h_{es},K_{n}\right),\left(h_{in},K_{in}\right) \right\}$, note that all elements in $\mathcal{I}$ are defined above in \textbf{situation 1} and \textbf{situation 2}, i.e., (\ref{PIctl}), (\ref{decrease_al}) to (\ref{decrease_limit2}) and (\ref{increase_al}) to (\ref{increase_last}). This is the general formulation of the previous situation, where specific situations are taken into consideration when certain conditions are met. Otherwise, the operational point is determined by comparing the three objective functions, i.e., $h/C(K)$. The algorithm for this VSOO-DT approach is outlined below.

\begin{algorithm}
\caption{The Variable-Step-Optimal-Order DT Method}\label{alg:VSOO-DT}
\begin{algorithmic}[1]
\REQUIRE ${t_0, }$ ${t_e, }$ ${h_0, }$ ${K_0, }$ ${\mathbf{x_0}, }$ ${Tol, }$ $h_{\min}, $ $h_{\max}, $ $K_{\min}, $ $K_{\max}, $
${\mu_{de}, }$ ${\mu_{in}, }$ ${\delta K, }$ ${\Delta K, }$ $\mathbf{\eta, }$ ${\gamma, }$ $\theta_{\max}. $

\STATE initialization $t=t_0, n=0,e_{0}=0,\mathbf{X}(0)=\mathbf{x_0}, \mathbf{Y}(0)=\mathbf{G}(\mathbf{X}(0)), \mathbf{Z}(0) =\left(\mathbf{X}^\top(0),\mathbf{Y}^\top(0)\right)^\top$

\WHILE{$t+h_n<t_e$}
    \WHILE{$k<K_n$}
        \STATE  $\mathbf{Y}(k) \gets \mathbf{G}(\mathbf{X}(k),\mathbf{X}(k-1) \cdots \mathbf{X}(0))$
        \STATE $\mathbf{Z}(k) \gets (\mathbf{X}(k),\mathbf{Y}(k))$ 
        \STATE $\mathbf{X}(k+1) \gets {(k+1)^{-1}}\mathbf{J}\mathbf{Z}(k)$
        \STATE $k \gets k+1$
    \ENDWHILE
\STATE $\mathbf{x}(t_{n+1}) \gets \sum^{K_n}_{k=0}{\mathbf{X}(k)h^k}$

\IF{$h_{es}\leq h_{\min}$}
    \STATE $h_{es}\gets h_{\min}$
\ELSIF{$h_{es}\geq h_{\max}$}
    \STATE $h_{es}\gets h_{\max}$
\ENDIF

\STATE  calculating step $h_{es}$  using PI control strategy $(\ref{PIctl})$
\IF{\color{black}$(\ref{decrease_condi})\text{ and }(\ref{increase_condi})$ \color{black}hold and $K_{\min}\leq K_{de}\leq K_{in}\leq K_{\max}$}
    \STATE   $ \left(h_{n+1},K_{n+1}\right)\gets \arg\max\limits_{\substack{\left(h,K\right)\in\mathcal{I}}}\frac{h}{C(K)}$
\ELSIF{$\eqref{decrease_condi},\eqref{decrease}$ hold and $K_{\min}\leq K_{de}$}
    \STATE $(h_{n+1},K_{n+1})\gets\left(h_{de},K_{de}\right)$ by $(\ref{decrease_al})-(\ref{decrease_limit2})$
\ELSIF{$\eqref{increase_condi},\eqref{increase}$ hold and $K_{\min}\leq K_{de}$}
    \STATE $(h_{n+1},K_{n+1})\gets\left(h_{in},K_{in}\right)$ by $(\ref{increase_al})-(\ref{increase_last})$
\ELSE
    \STATE $(h_{n+1},K_{n+1})\gets\left(h_{es},K_{n}\right)$ by $(\ref{PIctl})$
\ENDIF

\STATE ${{\mathbf{X(0)}} \gets {\mathbf{x}}_{n+1}}$
\STATE $t_{n+1} \gets t+h_n, t \gets t+h_n, n \gets n+1, k \gets 0$

\ENDWHILE
\RETURN $\mathbf{x}(t), $ $t_0\leq{t}\leq{t_e}$
\end{algorithmic}
\label{DT}
\end{algorithm}
\vspace{-2mm}
\subsection{Selecting Parameters}
With the basic motivations and principles of the PI control strategy and the optimization strategy explained, the selection of each parameter with the strategies is suggested below:

\textit{Initial Values}: Begin with a small initial step size ($h_0$) and order ($K_0$) to have a safe start. Notably, the proposed methods' robustness allows these values to adapt during the simulation, reducing sensitivity to initial settings.

\textit{Error Tolerance ($Tol$)}: Select the $Tol$ value based on specific simulation requirements, typically ranging from $10^{-25}$ to $10^{-2}$. Higher $Tol$ values are primarily used to assess performance but are generally unnecessary. Typically, a range of $10^{-8}$ to $10^{-2}$ is sufficient.

\textit{Additional Parameters}: Parameters such as minimum and maximum step sizes ($h_{\min}$ and $h_{\max}$) are problem-specific and designed for practical purposes, like achieving desired solution resolutions.  Minimum and maximum orders ($K_{\min}$ and $K_{\max}$) are flexible depending on the problem. Note that high orders are rarely needed unless exceptional accuracy is required.

\textit{Thresholds for Order Switching (${\mu_{de}}$ and $\mu_{in}$)}: These thresholds relate to the order-switching process within the VSOO-DT method. Values exceeding 1 impose stricter switching criteria, ensuring superior performance in the new state. Notably, $\mu_{in}$ should exhibit a normalizedly higher value compared to $\mu_{de}$ since $K_{in}^\text{th}$ order coefficients are estimated by $\rho_{in}$ but $K_{de}^\text{th}$ order coefficients could be computed accurately.

\textit{Order Adjustment Parameters ($\delta K$ and $\Delta K$)}: These parameters can be adjusted as needed. This paper lets $\delta K = \Delta K = 1$ to achieve a smooth transition between orders. Alternatively, order switching can occur after several steps with larger $\delta K$ and $\Delta K$ values.

\textit{Scaling Factor for Error Estimation ($\eta_i$)}: $\eta_i$ scales the estimated error. It helps to avoid singularities when certain variables approach zero. Setting $\eta_i = 0$ results in purely normalized error estimation, while $|\eta_i| > 0$ allows a balance between absolute and normalized errors. This point will become clear if setting $\eta_i= -\left|[\mathbf{x}]_i\right|+1$. Generally, the \eqref{error2} can capture the hybrid absolute-normalized error, and one can adjust the ratio between absolute and normalized errors, a small $|\eta_i|$ will lead to a domination of the normalized error which is usually meaningful and reliable for simulating a large-scale power system having state variables with different physical meaning. 

\textit{Safety Factors and Limits ($\gamma$ and $\theta_{max}$)}: $\gamma$ acts as a safety factor and can be set to 1 in the proposed VSOO-DT method due to the presence of a P controller. $\theta_{max}$ limits the maximum step size increment to prevent excessive overshooting, e.g., $\theta_{max}=2$ means that the largest $h_{n+1}$ will not exceed $2h_n$. In power system simulations, double the step size would be enough to respond to the dynamics of systems without influencing the resolution of solutions.

Values and ranges of these parameters are suggested in Table ~\ref{Parameters Setting} for power system simulations. It is worth noting that most of the parameters are adaptable to meet any specific requirements.

\begin{table}
    \centering
    {\color{black}
    \caption{Setting of Parameters}
    \label{Parameters Setting}
    
    \begin{tabular}{ccc}
    \toprule
         Parameters&  Meanings& Values\\
    \midrule
       ${h_0}$& initial step-size& 0.001\\
       $K_0$& initial order&8\\
       $Tol$& error tolerance&$\interval{10{^{-25}}}{10{^{-2}}}$\\
       $h_{\min}$& minimum step-size&depends\\
       $h_{\max}$& maximum step-size&depends\\
       $K_{\min}$& minimum order&$\geq4$  \\
       $K_{\max}$& maximum order&$\leq45$ \\
       ${\mu_{de}}$& switching factor for decreasing order & $\interval{1}{2.5}$\\
       ${\mu_{in}}$& switching factor for increasing order & $\interval{1}{2.5}$\\
       ${\delta K}$& increased order & $\geq1$\\
       ${\Delta K}$& decreased order & $\geq1$\\
       $\mathbf{\eta}_i$&  scaling factor for component $[\mathbf{x}]_i$& $\interval{10{^{-19}}}{0.1}$\\
       ${\gamma}$& safety factor & $\interval{0.85}{1}$\\
       $\theta_{\max}$ & maximum incremental gain & $\interval{1.25}{2}$\\
           \bottomrule
    \end{tabular}}
\end{table}
\color{black}
\section{Numerical Results}
In this section, the efficiency, robustness, and validity of the proposed algorithm are evaluated using both classical and detailed models of the IEEE 9-bus system, IEEE 39-bus system, and \color{black}{Polish} \color{black} 2383-bus system \cite{liu2019power}. The case study focuses on three stages: (1) verification of the proposed approach with different scenarios in different systems, (2) validation of the robustness with different tolerances and comparison of the proposed approach with different fixed orders using variable time step algorithm, (3) comparison of the efficiency of the proposed approach with other numerical algorithms. To compare the error of each method, the result from the RK4 method with a tiny time step size $h = 1 \times 10^{-4}$ s is considered the benchmark.
\subsection{Screening of Comprehensive Scenarios}
The proposed VSOO-DT method is tested on three systems: IEEE 9-bus, IEEE 39-bus, and \color{black} Polish \color{black} 2383-bus \color{black}systems\color{black}. Simulations are conducted using Matlab on a laptop computer with an Intel CoreTM i7-6600U CPU and 8 G\color{black}B \color{black} RAM. For the IEEE 9-bus system, simulations are conducted with and without the tripping of the branch between buses 5 and 7 after grounding. The parameters used in the simulations are provided in Table~\ref{p3_9}.\par
\begin{figure}[!ht]
\vspace{-4mm}
\color{black}
\centering
\subfloat[]{\includegraphics[width=1.8in]{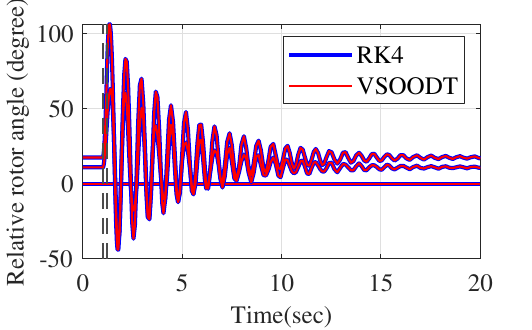}%
\label{ag3_9}}
\hfil\hspace{-4mm}
\subfloat[]{\includegraphics[width=1.8in]{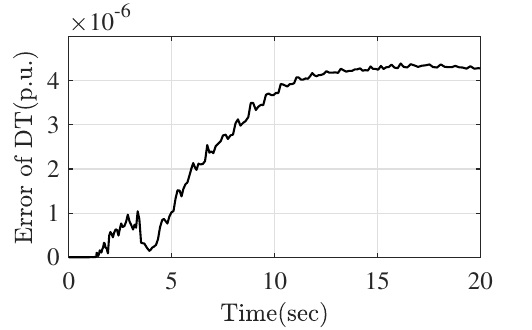}%
\label{er3_9}}
\vspace{-4mm}
\hfil\hspace{-4mm}
\subfloat[]{\includegraphics[width=1.8in]{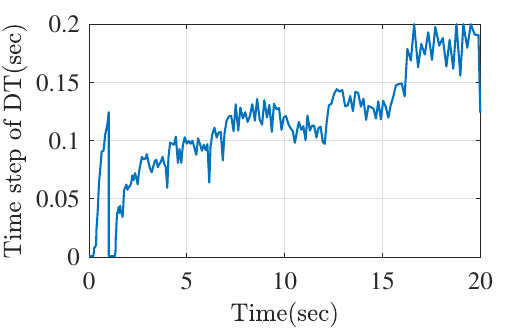}%
\label{stp3_9}}
\hfil\hspace{-4mm}
\subfloat[]{\includegraphics[width=1.8in]{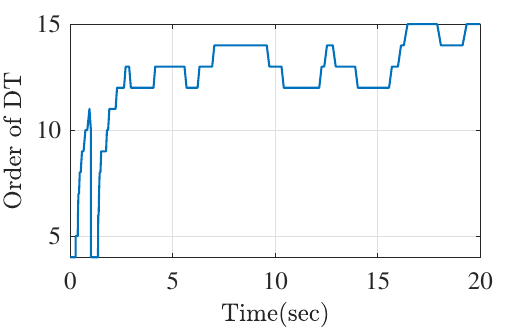}%
\label{od3_9}}\\
\vspace{-4mm}
\subfloat[]{\includegraphics[width=1.8in]{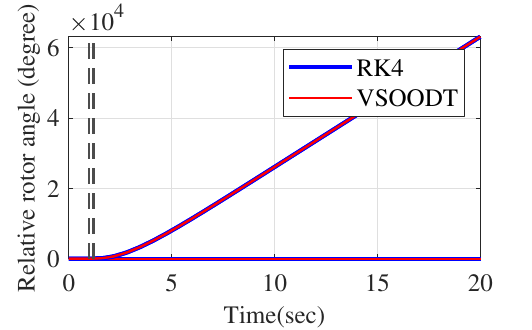}%
\label{ag3_9u}}
\hfil\hspace{-4mm}
\subfloat[]{\includegraphics[width=1.8in]{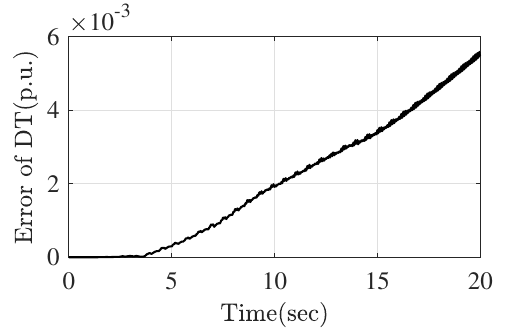}%
\label{er3_9u}}
\vspace{-4mm}
\hfil\hspace{-4mm}
\subfloat[]{\includegraphics[width=1.8in]{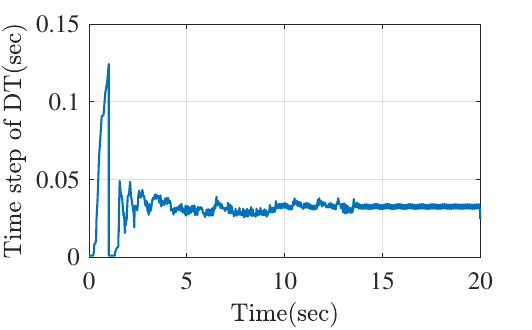}%
\label{stp3_9u}}
\hfil\hspace{-4mm}
\subfloat[]{\includegraphics[width=1.8in]{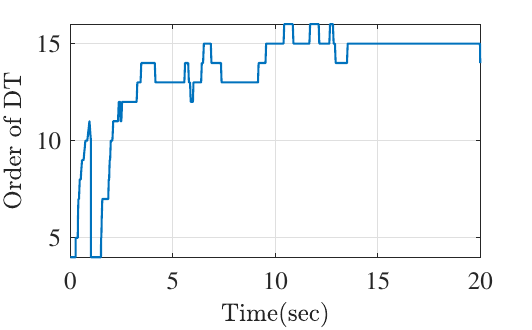}%
\label{od3_9u}}
\vspace{-2mm}
\caption{Simulation results for the IEEE 9-bus system. (a) Relative rotor angles in the stable case. (b) maximum errors in the stable case. (c) Step size during the simulation in the stable case. (d) Optimal order of DT during the simulation in the stable case. (e) Relative rotor angles in the unstable case. (f) maximum errors in the unstable case. (g) Step size during the simulation in the unstable case. (h) Optimal order of DT during the simulation in the unstable case.}
\label{fig_3bus}
\end{figure}

\begin{table}[!ht]
\begin{center}
\caption{Default Setting of Parameters for IEEE 9-Bus System}
\label{p3_9}
\begin{tabular}{c  c  c  c  c  c  c }
 \toprule
${h_0}$ & $K_0$ & $Tol$     & $h_{\min}$ &  $h_{\max}$ & $K_{\min}$ & $K_{\max}$\\
\midrule
0.001 s   & 4     & 10$^{-5}$ & 10$^{-4}$ s &  0.2 s       &  4         &    45  \\
\toprule
${\mu_{de}}$ & ${\mu_{in}}$ & ${\delta K}$ & ${\Delta K}$ & $\mathbf{\eta}$& ${\gamma}$ & $\theta_{\max}$\\
\midrule
1   & 1.02     & 1 & 1 &  10$^{-19}$        &  1         &    2  \\
\bottomrule
\end{tabular}
\end{center}
\end{table}
\color{black}
The simulation results of several variables, including relative rotor angles, absolute error, adaptive step size, and the optimal order during the simulation, are displayed in Fig. \ref{ag3_9}-\ref{od3_9u}. The sequence of error in Fig. \ref{er3_9} is calculated by the infinity norm of the difference between the benchmark $\mathbf{x_{bh}}(t_n)$ and $\mathbf{x_n}$ for every step, i.e., given a time $t_n$, the error is calculated:
\begin{equation}
    \text{Error}(t_n) = \lVert \mathbf{x_{bh}}(t_n)-\mathbf{x_n} \rVert_\infty.
\end{equation}
From Fig. \ref{er3_9}, the error is finally stabilized around $4\times 10^{-6}$ which means the accuracy of the proposed method can be guaranteed. The step size and order are time-varying and controlled by the PI controller and the optimal order algorithm. By increasing the order during the simulation, a relatively large time step size can be applied without violating the numerical stability as well as with promising accuracy. Note that the step size finally is approaching the upper limit $h_{\max}=0.2~\mathrm{s}$ and the order is also stabilized around 15. For the unstable case, the step size can be stabilized around 0.04 s and the order oscillates around 15, since compared with the stable one, the system is stiffer, thus requiring a more minor time step. \par

\begin{figure}[!ht]
\vspace{-4mm}
\color{black}
\centering
\subfloat[]{\includegraphics[width=1.8in]{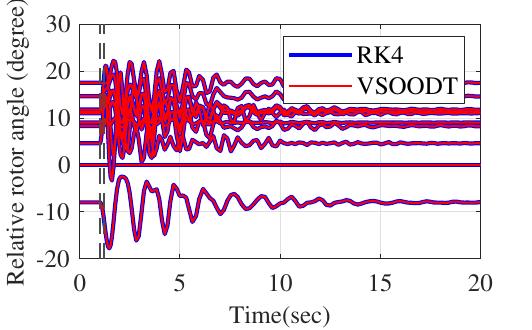}%
\label{ag39}}
\hfil\hspace{-4mm}
\subfloat[]{\includegraphics[width=1.8in]{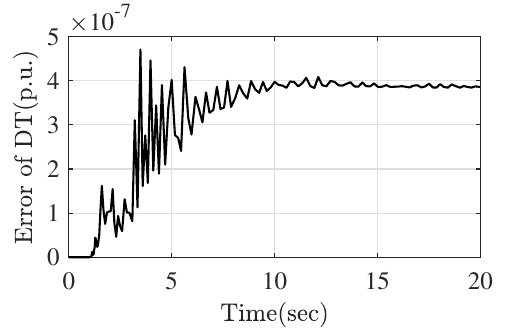}%
\label{er39}}
\vspace{-4mm}
\hfil\hspace{-4mm}
\subfloat[]{\includegraphics[width=1.8in]{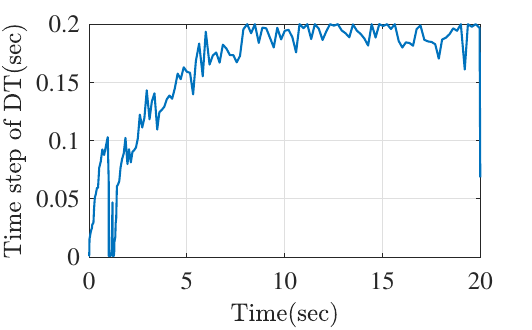}%
\label{stp39}}
\hfil\hspace{-4mm}
\subfloat[]{\includegraphics[width=1.8in]{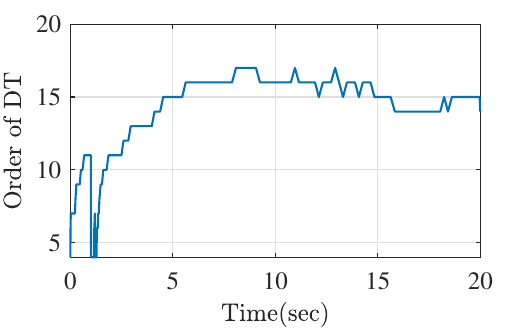}%
\label{od39}}\\
\vspace{-4mm}
\subfloat[]{\includegraphics[width=1.8in]{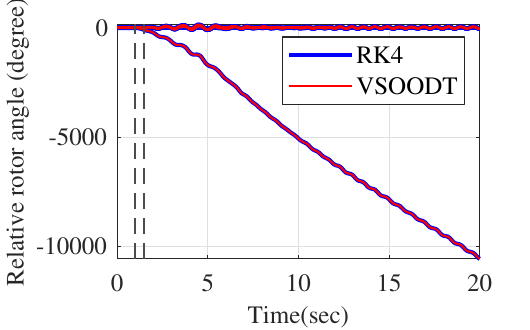}%
\label{ag39u}}
\hfil\hspace{-4mm}
\subfloat[]{\includegraphics[width=1.8in]{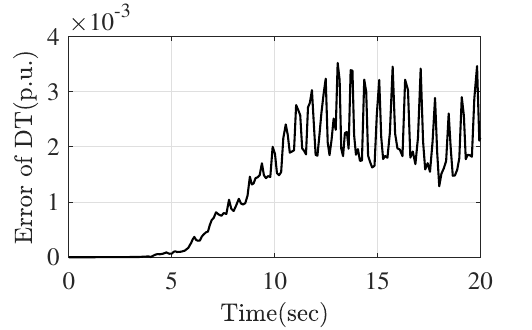}%
\label{er39u}}
\vspace{-4mm}
\hfil\hspace{-4mm}
\subfloat[]{\includegraphics[width=1.8in]{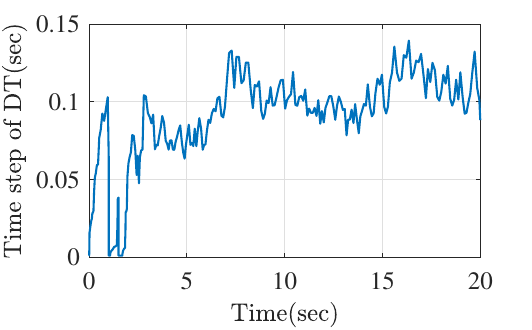}%
\label{stp39u}}
\hfil\hspace{-4mm}
\subfloat[]{\includegraphics[width=1.8in]{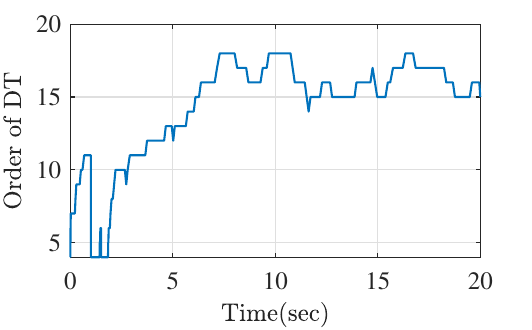}%
\label{od39u}}
\vspace{-2mm}
\caption{Simulation results for the IEEE 39-bus system. (a) Relative rotor angles in the stable case. (b) maximum errors in the stable case. (c) Step size during the simulation in the stable case. (d) Optimal order of DT during the simulation in the stable case. (e) Relative rotor angles in the unstable case. (f) maximum errors in the unstable case. (g) Step size during the simulation in the unstable case. (h) Optimal order of DT during the simulation in the unstable case.}
\label{fig_39bus}
\end{figure}

For the IEEE 39-bus system, a three-phase bus grounding fault is applied to bus 9 at 1 s and cleared at 1.2 s, resulting in potential rotor angle instability. An unstable case is considered where a three-phase bus grounding is applied to bus 39 at 1 s and cleared at 1.5 s by tripping the generator at bus 39. The same parameters as in Table~\ref{p3_9} are used. Simulation results are displayed in Fig. \ref{ag39}-\ref{od39u}. In this case, initially \textbf{situation 2} is the dominant mode, and the step size increases rapidly with the increasing order. The upper limit of the step size is approached around 10s, and the order also stops to increase as there is no further room to increase the step size for acceleration. \textbf{Situation 1 or 3} is activated, and such a high-order model does not need to be maintained. Nonetheless, errors can still be controlled. For the unstable case, the step size is increased and oscillates around 0.1 s, and the order is adjusted to increase. Otherwise, the time step size will start to decrease to influence the time performance.\par

\begin{figure}[!ht]
\color{black}
\vspace{-4mm}
\centering
\subfloat[]{\includegraphics[width=1.8in]{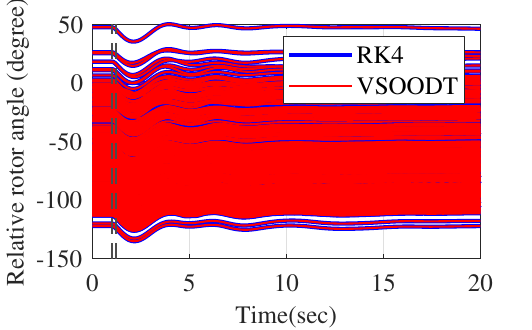}%
\label{ag2383}}
\hfil
\hspace{-4mm}
\subfloat[]{\includegraphics[width=1.8in]{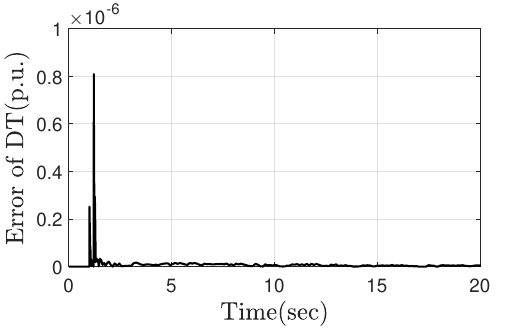}%
\label{er2383}}
\vspace{-4mm}
\hfil
\hspace{-4mm}
\subfloat[]{\includegraphics[width=1.8in]{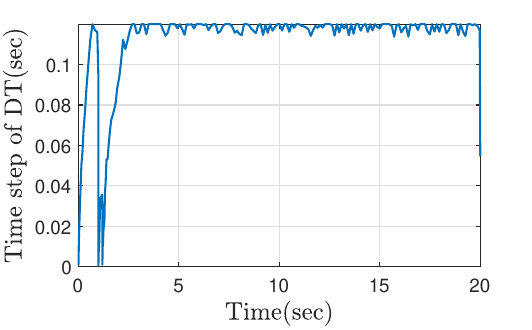}%
\label{stp2383}}
\hfil
\hspace{-4mm}
\subfloat[]{\includegraphics[width=1.8in]{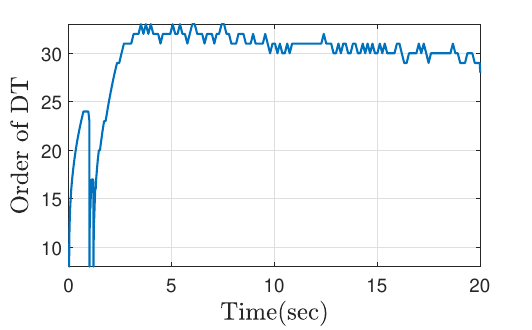}%
\label{od2383}}\\
\vspace{-4mm}
\subfloat[]{\includegraphics[width=1.8in]{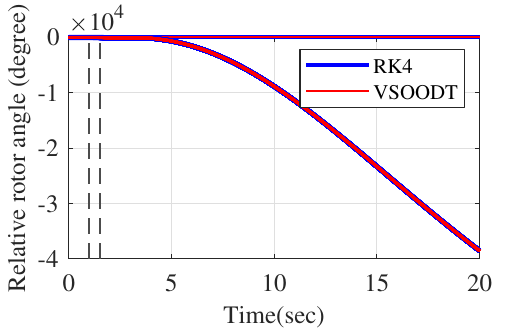}%
\label{ag2383u}}
\hfil
\hspace{-4mm}
\subfloat[]{\includegraphics[width=1.8in]{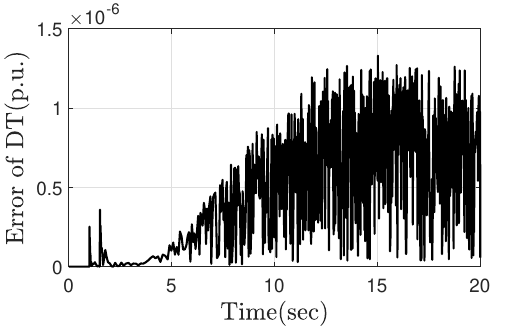}%
\label{er2383u}}
\vspace{-4mm}
\hfil
\hspace{-4mm}
\subfloat[]{\includegraphics[width=1.8in]{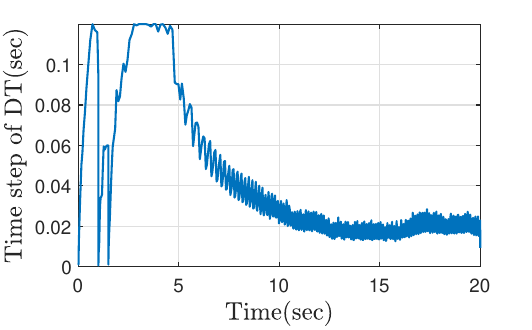}%
\label{stp2383u}}
\hfil
\hspace{-4mm}
\subfloat[]{\includegraphics[width=1.8in]{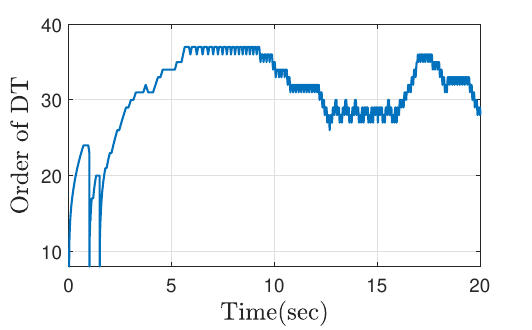}%
\label{od2383u}}
\vspace{-2mm}
\caption{Simulation results for the Polish 2383-bus system. (a) Relative rotor angles in the stable case. (b) maximum errors in the stable case. (c) Step size during the simulation in the stable case. (d) Optimal order of DT during the simulation in the stable case. (e) Relative rotor angles in the unstable case. (f) maximum errors in the unstable case. (g) Step size during the simulation in the unstable case. (h) Optimal order of DT during the simulation in the unstable case. }
\label{fig_2383bus}
\end{figure}

Lastly, the Polish 2383-bus system is considered, where a three-phase bus grounding is applied to bus 9 at 1 s and the branch between bus 6 and bus 9 is tripped at 1.2 s, leading to a stable system. An unstable case arises from tripping the branch between bus 6 and bus 9 at 1.5 s. The parameters are shown in Table~\ref{p3_9} for the above two cases except for $K_0=8$ and $h_{\max}=0.12 s$. Simulation results for $Tol=10^{-15}$ are displayed in Fig. \ref{fig_2383bus}. In the stable case, the step size approaches the upper limit at around 3 s with the order increasing up to 32. The system gradually converges to an equilibrium point with the order ultimately converging to 29. Note that the order will keep decreasing if the simulation continues beyond 20 s since a lower-order model can support the $h_n$ approach to the upper limit. The unstable case has the step size approach the upper limit before one rotor angle diverges after 5 s, and reduces the step size rapidly due to the faster dyanmics than the stable case. The order with the unstable case is lower, and increasing the order does not significantly improve the simulation speed. Overall, the proposed VSOO-DT method demonstrates its robustness in different cases with the desirable numerical stability and enlarged step sizes. \par

\color{black}
\subsection{Comparing with the Variable-Step DT Method with a Fixed Order}
\color{black}
To emphasize the importance of selecting the optimal order, the proposed VSOO-DT method is compared with the proposed VS-DT method using different fixed orders. The parameter settings follow the last subsection and those stable cases are considered. System dynamics within 20 s are simulated, and the corresponding computational times are compared under different tolerances ($Tol$), as shown in Table~\ref{table:2c}. 

When $Tol=10^{-15}$, an increase of the order leads to a longer execution time for the 9-bus and 39-bus systems since the increased complexity by using a higher order offsets the benefit from using a larger time step size. For these cases, the optimal order is 15. For the 2383-bus system, the optimal order is 25. Note that the execution time decreases when the order changes from 15 to 25 but increases after the order of 25. This highlights the benefit of a higher-order solution for simulating a large-scale system, allowing for a larger step size to accelerate the simulation. However, the benefit may decrease and vanish beyond a certain order.

Similar situations occur in the 9-bus and 39-bus systems when $Tol$ is increased to $10^{-20}$. 25 is the ideal order with minimum execution time. However, for the 2383-bus system, the minimum execution time is achieved at order 35. When $Tol$ approaches $10^{-25}$, the ideal order for the 9-bus system is 45, but for the 39-bus system, it is 35. This is because the lower dimension of the 9-bus system allows for employing a higher order VS-DT method, with a larger time step size, which outperforms the use of a lower order model with a smaller step size. For large-scale systems, the optimal order of the VS-DT method remains 35 when $Tol= 10^{-25}$, indicating that simply increasing the order does not always help reduce computational time. While the VS-DT method provides promising results across different models and tolerances, the determination of a universally ideal order remains unsolved. However, the results of the VSOO-DT method outperform the VS-DT method under all tolerances because of the additional optimal order control strategy. These findings affirm the robustness and efficiency of the proposed optimal order strategy in comparison to the VS-DT method.\par
\begin{figure}[!ht]
\vspace{-4mm}
\color{black}
\centering
\subfloat[]{\includegraphics[width=0.5\columnwidth]{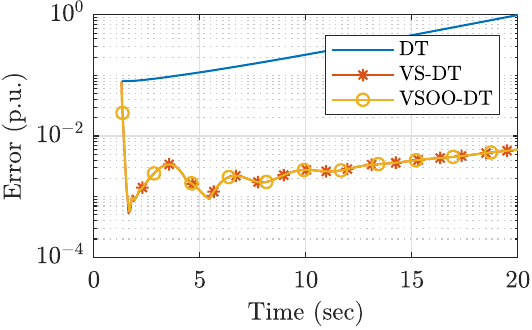}%
\label{stablility_error}}
\subfloat[]{\includegraphics[width=0.5\columnwidth]{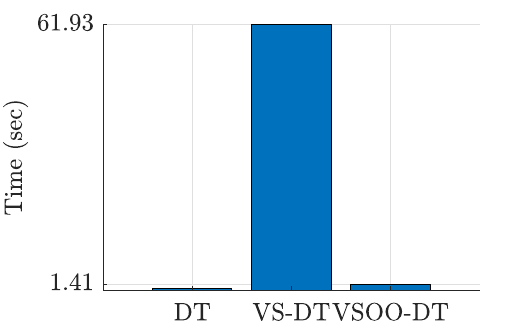}%
\label{stablility_step}}
\caption{Stability comparison for the DT, VS-DT, and VSOO-DT methods for the 2383-bus system. (a)Errors comparison. (b)Execution time comparison.}
\label{stability}
\end{figure}

\color{black}
\begin{table}[!ht]
\vspace{-4mm}
\begin{center}
\setlength{\tabcolsep}{0.1pt}
\caption{Computational Time with Different Fixed Order VS-DT Methods}
\label{table:2c}
     \begin{tabular}{c c c c c}
        \toprule
\thead {Order\\$K$}  
    & {\thead{Tolerance \\$Tol$} }
        & {\thead{CPU  Time(s)\\9-bus }} 
        & {\thead{CPU  Time(s)\\39-bus }}
        & {\thead{CPU  Time(s)\\2383-bus }} \\ [0.5ex]
        \midrule
15        & $10^{-15}$ & 0.2887 & 0.3356 & 18.9236\\
            & $10^{-20}$ & 0.7169 & 0.6177 & 45.3932 \\
            & $10^{-25}$ & 1.5774 & 1.5614  & 110.2162 \\ 
25        & $10^{-15}$ & 0.3336 &0.3629 & 10.4229\\
            & $10^{-20}$ & 0.5263 & 0.4411 & 20.9032\\
            & $10^{-25}$ & 0.8235 & 0.7110 & 30.4770\\ 
35        & $10^{-15}$ & 0.4468 & 0.5695 & 14.4394\\
            & $10^{-20}$ & 0.5981 & 0.5836 & 17.7867\\
            & $10^{-25}$ & 0.7329 & 0.6255 & 22.4509\\ 
45        & $10^{-15}$ & 0.5680 & 0.7359 & 24.5646\\
            & $10^{-20}$ & 0.6138 & 0.8080 & 25.6463\\
            & $10^{-25}$ & 0.7029 & 0.7811 & 26.6313\\ 
Optimal Order  & $10^{-15}$  & \B 0.3712 & \B 0.3291 & \B 10.1469\\
\color{black}{(Start from $K_0=8$)}\color{black}          & $10^{-20}$  & \B0.6047   & \B0.4976  & \B14.5964\\
                            & $10^{-25}$  & \B0.8858   & \B0.5557  & \B21.0387\\
        \bottomrule
    \end{tabular}
    \end{center}
    \vspace{-2mm}
\end{table}
\color{black}
To further compare the numerical stability and speed of the DT, VS-DT, and VSOODT methods, consider a three-phase grounding fault applied to bus 3 and cleared after 0.2s for the 2383-bus system. Parameters in Table~\ref{p3_9} are utilized without setting $h_{max}$. A large step size $h_0=0.1\mathrm{~s}$ is employed and the error is shown in Fig.~\ref{stablility_error}. The errors of the VS-DT and VSOO-DT methods can be maintained at less than $10^{-2}$ but the DT method is diverged. It can be concluded that both VS-DT and VSOO-DT methods improve the numerical stability compared with the DT method. 
Furthermore, results of the execution time are shown in Fig.~\ref{stablility_step}. While the VS-DT method is numerically stable under a large step size, its order $K=4$ still limits the computational performance. The proposed VSOO-DT method achieves the same level of accuracy with a much less execution time (over 40 times speed up) because of the proposed optimal order strategy. In summary, the failure of the original DT method due to the improper step size and order can be partially solved by the VS-DT method and fully addressed by the VSOO-DT method.
\color{black}
\subsection{Comparison with Classical Numerical Method}
\begin{figure}[!ht]
\vspace{-5mm}
\color{black}
\centering
\includegraphics[width=0.85\columnwidth]{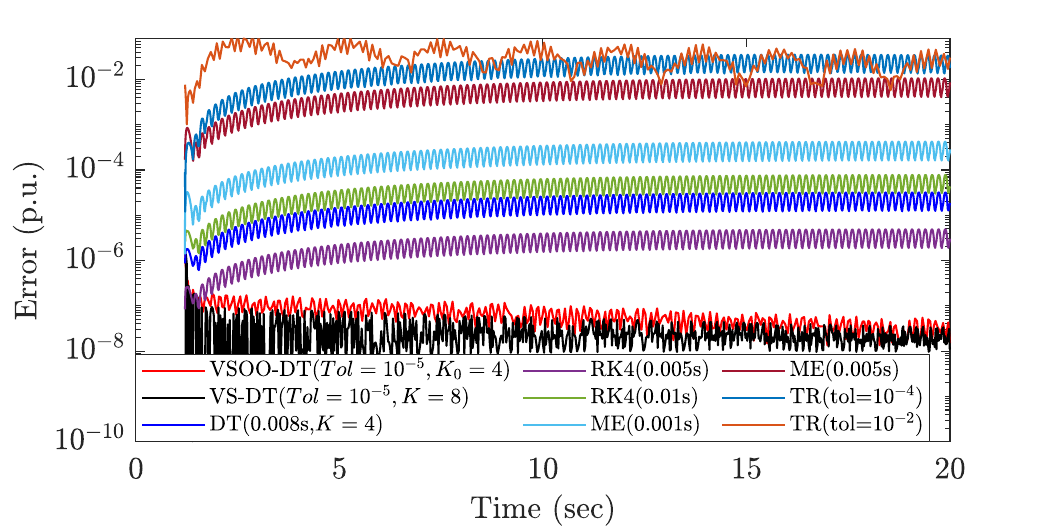}
\vspace{-4mm}
\label{errorcomp}
\caption{Simulation errors of the 2383-bus system using different methods}
\label{2383different}
\end{figure}

The proposed method is compared with the RK4 method which is applied in the power system simulation on three systems under the same error level. For the 9-bus system and 39-bus system, previous stable cases are still considered. The results are shown below in Table~\ref{table:rk4}, it can be observed the time performances of the VSOO-DT method are much better than those of the RK4 method under all different levels of errors.
\begin{table}[!ht]
\vspace{-2mm}
\begin{center}
\setlength{\tabcolsep}{10pt}
\caption{Computational Times with the RK4 and VSOO-DT Methods}

\label{table:rk4}
     \begin{tabular}{c c c c c}
        \toprule
\thead {System }  
    & {\thead{Error} }
        & {\thead{CPU Time(s)\\(RK4) }}  & {\thead{CPU Time(s)\\(VSOO-DT) }} \\ [0.5ex]
        \midrule
9-bus   & $10^{-3}$ & 0.2303 & 0.0812 \\
        & $10^{-4}$ & 0.4377 & 0.0933 \\
        & $10^{-5}$ & 0.5586 & 0.1078\\ 
39-bus  & $10^{-3}$ & 0.1469 & 0.0922\\
        & $10^{-4}$ & 0.3041 & 0.1410\\
        & $10^{-5}$ & 0.4038 & 0.1564\\
        \bottomrule
    \end{tabular}
    \end{center}
    \vspace{-2mm}
\end{table}
\color{black}
{A more comprehensive case involving a large disturbance for the 2383-bus system is studied. A three-phase grounding fault is applied to bus 1681 and then is cleared after 0.2 s. The simulation covers the system's dynamics over 20 seconds, and the resulting errors and computational times are compared across various numerical methods, including the RK4 method, the Modified Euler (ME) method, and the Trapezoidal Rule (TR) method. Note that the TR method refers to the optimized and superior MATLAB built-in solver (ode23t), which also incorporates the variable step strategy. Table~\ref{table:errorcomp} provides the comparison of the execution times and mean-max errors, here "mean" errors represent average errors for maximum errors along the time. Fig~\ref{2383different} displays the maximum errors during the post-fault stage for different methods. The proposed VS-DT and VSOO-DT methods exhibit a significant advantage in terms of both efficiency and accuracy when compared to RK4, ME, and TR methods. In particular, to compare with the conventional DT method, both DT and VSOO-DT methods start from identical conditions, i.e., $K=K_0=4$, $h=h_0=0.008s$ and $\mu_{in}=1.5$, and other parameters are shown in Table~\ref{p3_9}. The results show that the VSOO-DT method is faster than the DT and VS-DT methods and it has the same level of high accuracy compared to the VS-DT method. The proposed VSOO-DT method effectively adjusts the order and step size on the fly of the simulation, affirming the efficiency and importance of the optimal order strategy. Moreover, the performance of the DT highly depends on the selection of the initial step size and order, which requires extra experience. This limitation has been verified in the previous stability study. In cases where the settings are not ideal, the execution time can become prolonged, or the accuracy may not be promising.

In summary, the proposed VSOO-DT method not only attains high accuracy like the VS-DT method but also has the fastest computational speed among all other methods even when the initial settings are not ideal.
}\color{black}



\begin{figure}[!ht]
\vspace{-4mm}
\color{black}
\centering
\includegraphics[width=0.65\columnwidth]{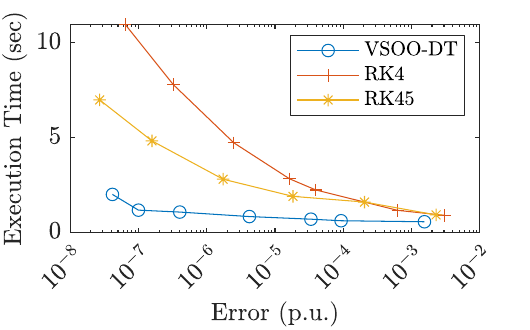}
\vspace{-2mm}
\caption{Comparison of execution times for different errors by three methods}
\label{Error_rk45}
\vspace{-3mm}
\end{figure}

\begin{figure}[!ht]
\vspace{-5mm}
\color{black}
\centering
\subfloat[]{\includegraphics[width=1.8 in]{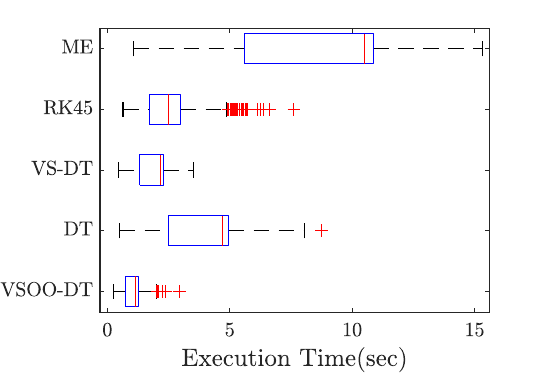}%
\label{box}}\hspace{-4mm}
\subfloat[]{\includegraphics[width=1.8 in]{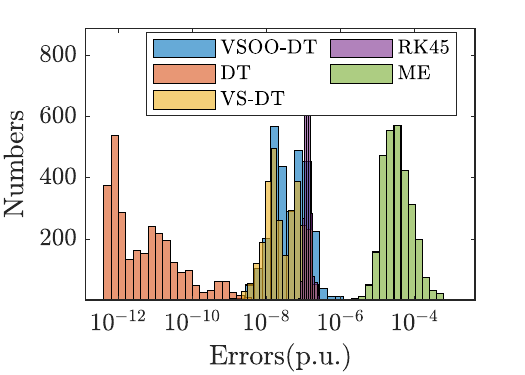}%
\label{his}}
\caption{Distribution of N-1 dynamic screenings. (a)Execution times distributions. (b)Mean-Max error distributions.}
\label{N-1}
\vspace{-3mm}
\end{figure}

\begin{table}[!ht]
\vspace{-2mm}
\begin{center}
\setlength{\tabcolsep}{0pt}
\color{black}
\caption{Performances of Different Methods on the 2383-bus System}
\label{table:errorcomp}
\begin{tabular}{lcc}
        \toprule
 Method & CPU time $(\mathrm{s})$ & Mean-Max Error(p.u.) \\
\midrule
VSOO-DT$\left(Tol=10^{-5},K_0=4\right)$ & 1.8943  & 8.79$\times10^{-8}$ \\
VS-DT$\left(Tol=10^{-5},K=8\right)$ & 3.3317      & 3.11$\times10^{-8}$ \\
DT$\left(h=0.008 \mathrm{~s},K=4\right)$ & 3.7092 & 1.63$\times10^{-5}$ \\

RK4$(h=0.005 \mathrm{~s})$ & 4.3180 & 2.49$\times10^{-6}$ \\
RK4$(h=0.01 \mathrm{~s})$ & 2.2454 & 3.80$\times10^{-5}$ \\

ME $(h=0.001 \mathrm{~s})$ & 11.7445 & 2.15$\times10^{-4}$ \\
ME $(h=0.005 \mathrm{~s})$ & 2.4866 & 0.0054 \\
TR $(tol=10^{-4})$ & 146.8446 & 0.017 \\
TR $(tol=10^{-2})$ & 44.8635 & 0.033 \\
\bottomrule
\end{tabular}
    \end{center}
\end{table}
\color{black}
While the TR and ME methods cannot simultaneously match the accuracy and speed of the VSOO-DT method, other traditional methods like the RK4 method require an extremely small time step size, like 0.001 s, to achieve a comparable error level to the VSOO-DT method, but it significantly slows down the computational speed. To provide a more comprehensive comparison, a MATLAB built-in variable step solver (ode45), i.e., the RK45 method is also considered, which is known for its success and robustness.
By adjusting the tolerances for both VSOO-DT and RK45 methods, and the step sizes for the RK4 method, execution times across various levels of mean-max errors are compared in Fig~\ref{Error_rk45}. The results clearly demonstrate that the VSOO-DT method outperforms the RK4 and RK45 methods under different error requirements. Its impressive efficiency and scalability are further indicated by the nearly flat slope observed in the curve of the VSOO-DT method. Slopes of these characteristic curves represent the increased computational effort for increasing the accuracy. The flat slope of the VSOO-DT method verifies its adaptability through flexible order and step adjustments.
\color{black}
\color{black}
To demonstrate the robustness and efficiency of the proposed VSOO-DT method, a dynamic N-1 stability assessment is studied. All three-phase grounding faults in the Polish 2383-bus system are considered, and each fault lasts 0.2 seconds. The default simulation time is 20 s. Simulations would be terminated if the maximum relative rotor angle exceeds 1000 degrees, indicating instability of the system. The comparison between the DT, VS-DT, VSOO-DT, RK45 (tolerance set at $10^{-8}$), and ME (with a step size of 0.001s) methods is shown in Fig~\ref{N-1}. The N-1 stability assessment includes all 2898 contingencies which can effectively verify the scalability and robustness of the proposed methods.

The initial orders of the DT, VS-DT, and VSOO-DT methods are 8 for a fair comparison. The step size of the DT is 0.01 s to prevent the numerical collapse. Despite the DT's capacity for excessive accuracy, it is unnecessarily time-consuming. This aligns with the observation that the DT requires tiny step sizes to maintain numerical stability, resulting in time wastage. Note that all methods based on the variable step strategy achieve faster speeds. However, a fixed optimal order and step size for the DT method may exist, but there is no general way to find it. And this is the motivation to design the proposed VSOO-DT method. Notably, the VSOO-DT method achieves a remarkable 10-fold speedup compared to the ME and a 2-fold speedup compared to the RK45 method, while the latter is the fastest conventional numerical method with promising numerical stability. Importantly, the execution times of the VSOO-DT method exhibit even less variation than those of the RK45 method, all while maintaining the same accuracy level (lower than $10^{-7}$). 
This N-1 study validates the importance of generalizing the SAS method with the variable step size and order strategies in large-scale simulations. Based on the VSOO-DT method, commercial software tools can be developed with improved accuracy and numerical stability of simulation for efficient dynamic stability assessment.
\section{Conclusion}
This paper addresses a critical gap in power system dynamic simulations, where conventional methods often rely on the fixed step size and order, lacking a systematic approach to optimize these parameters on the fly of the simulation. While the DT method has demonstrated the potential for increased step sizes using SAS, the theoretical studies of the optimal relationship between the step size and order remain understudied.

Two innovative strategies are proposed to solve this challenge: the VS-DT and the VSOO-DT methods. The VS-DT introduces an automatic PI-controlled variable step strategy, leveraging a fixed order SAS to achieve high-speed performance without compromising accuracy, even for large-scale power systems. Building upon this, the VSOO-DT method optimally varies the order of the SAS, efficiently determining the optimal order for each step through a formulated optimization problem based on the complexity of the system model.

The key contributions of this paper are summarized. First, the VS-DT and VSOO-DT methods provide a dynamic framework for simulations, offering an optimal balance between numerical stability and simulation speed. The automatic adjustment of step size and the determination of the optimal SAS order are achieved on the fly of the simulation, ensuring efficiency and accuracy. Additionally, the proposed strategies are theoretically analyzed for numerical stability, offering insights into the selection of parameters.

Furthermore, case studies demonstrate the effectiveness of the proposed methods. Compared to existing approaches, the VS-DT and VSOO-DT methods exhibit robust and efficient simulations by adapting step size and optimizing the order as needed. Importantly, these strategies present a general framework applicable to the DT method and other semi-analytical or numerical methods with adjustable orders. This generality opens avenues for broader applications in power system simulations.

In summary, the proposed strategies represent a fundamental change in dynamic simulations, adapting the step size and optimizing the order for improved performance in different situations. The framework presented goes beyond the DT method, laying the foundation for progress in a wide range of simulation methods.
\color{black}
\appendix
Consider the detailed power system dynamic model depicted below, which includes several controllers:
\begin{subequations}\label{equ:1}
\begin{align}
T_{sv}\dot {p}_{sv} &= -{p}_{sv}+p_{ref}-\frac{1}{R} s_m, \label{equ:1a}\\
T_{ch}\dot {p}_{m} &= -{p}_{m}+p_{sv}, \label{equ:1b}\\
T_{\omega}\dot {v}_1 &= -({K}_{\omega}+K_{p}p_e+K_{v}v_t+v_1), \label{equ:1c}\\
T_{e}\dot {e}_{fd} &= {v}_{r}-K_{e}e_{fd}, \label{equ:1d}\\
T_{f}\dot {v}_{f} &= -{v}_{f}+K_{f}e_{fd}, \label{equ:1e}\\
T_{r}\dot {v}_{ts} &= -{v}_{ts}+v_{t}, \label{equ:1f}\\
T_{a}\dot {v}_{r} &=
\begin{cases}
-{v}_{r}+K_{a}(v_{ref}+v_s-v_{ts}-v_f),\\
{\text{ if}}\ v_{r\text{min}}<v_r<v_{r\text{max}}\\
{0,}{\text{ if}}\ v_{r\text{min}}=v_r,\dot{v}_r<0{\text{ or }}v_{r\text{max}}=v_r,\dot{v}_r>0, \label{equ:1g}
\end{cases}
\end{align}
\end{subequations}
where $v_s$ is defined as $K_{\omega}+K_{p}p_e+K_{v}v_t+v_1$ and  (\ref{equ:1a}) and (\ref{equ:1b}) represent the governor and turbine models, respectively. The PSS IEE\color{black}{E} \color{black}Type I model is given in  (\ref{equ:1c}), while the simplified  IEEE Type I exciter model is represented by  (\ref{equ:1d}) through (\ref{equ:1g}). Furthermore, for the synchronous machine, a detailed $6^{\text{th}}$ order model is considered:
\begin{equation}
\begin{split}\label{detailed}
\dot{\delta}&=\omega_s s_m, \\
2H\dot{s}_m &= p_m-v_di_d-v_qi_q-Ds_m, \\
T'_{q0}\dot{e}'_{d} &= -\frac{x_q-x''_q}{x_q'-x''_q}e'_d+\frac{x_q-x'_q}{x_q'-x''_q}e''_d, \\
T'_{d0}\dot{e}'_{q} &= -\frac{x_d-x''_d}{x_d'-x''_d}e'_q+\frac{x_d-x'_d}{x_d'-x''_d}e''_q+e_{fd},\\
T''_{q0}\dot{e}''_{d} &= e'_d-e''_d+(x_q'-x''_q)i_q, \\
T''_{d0}\dot{e}''_{q} &= e'_q-e''_q-(x_d'-x''_d)i_d,     
\end{split}
\end{equation}
where $i_{xy}=\mathbf{Y}_{\mathbf{r}} v_{xy}$ and
\begin{equation}
\begin{gathered}
{\left[\begin{array}{l}
i_d \\
i_q
\end{array}\right]=\left[\begin{array}{cc}
r_a & -x_q^{\prime \prime} \\
x_d^{\prime \prime} & r_a
\end{array}\right]^{-1}\left(\left[\begin{array}{c}
e_d^{\prime} \\
e_q^{\prime}
\end{array}\right]-\left[\begin{array}{c}
v_d \\
v_q
\end{array}\right]\right)}, \\
{\left[\begin{array}{l}
i_x \\
i_y
\end{array}\right]=\mathbf{R}\left[\begin{array}{l}
i_d \\
i_q
\end{array}\right],\left[\begin{array}{l}
v_x \\
v_y
\end{array}\right]=\mathbf{R}\left[\begin{array}{l}
v_d \\
v_q
\end{array}\right]},
\end{gathered}
\end{equation}
and 
\begin{equation}
 \mathbf{R}=\left[\begin{array}{cc}
\sin \delta & \cos \delta \\
-\cos \delta & \sin \delta
\end{array}\right]. 
\end{equation}
For a more in-depth understanding of (\ref{equ:1}) and (\ref{detailed}), please refer to \cite{liu2019power}.

\bibliographystyle{IEEEtran}

\input{VSOODT.bbl}


\begin{IEEEbiography}[{\includegraphics[width=1in,height=1.25in,clip,keepaspectratio]{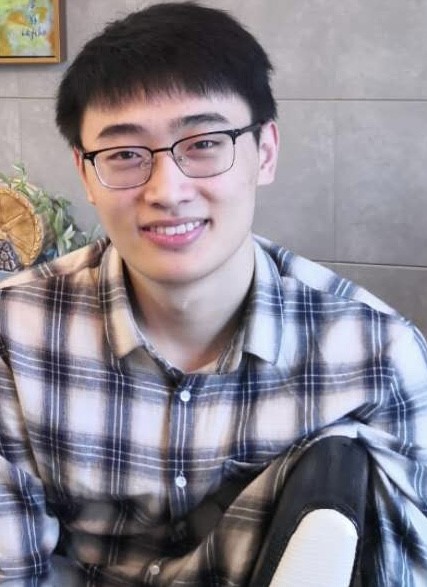}}]{Kaiyang Huang}
(Student Membe, IEEE) received the
B.S. degree in electrical engineering
from North China Electric Power University, China, in 2020.
He is currently pursuing the Ph.D. degree at the
Department of Electrical Engineering and Computer
Science, University of Tennessee, Knoxville, USA.
His research interests include power system simulation, transient
stability analysis, and dynamics.  \end{IEEEbiography}

\begin{IEEEbiography}[{\includegraphics[width=1in,height=1.25in,clip,keepaspectratio]{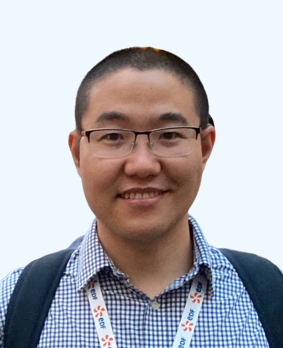}}]{Yang Liu}
(Member, IEEE) received the B.S.
degree in energy and power engineering from
Xi’an Jiaotong University, China, in 2013, the
M.S. degree in power engineering from Tsinghua
University, China, in 2016, and the Ph.D. degree
in electrical engineering from the University of
Tennessee, Knoxville, USA, in 2021. He was a
Post-Doctoral Research Associate with the University of Tennessee, in 2022. Currently, he is
a Post-Doctoral Appointee with the Argonne
National Laboratory. His research interests include stability, dynamics, and control of power grids and other complex systems. \end{IEEEbiography}

\begin{IEEEbiography}[{\includegraphics[width=1in,height=1.25in,clip,keepaspectratio]{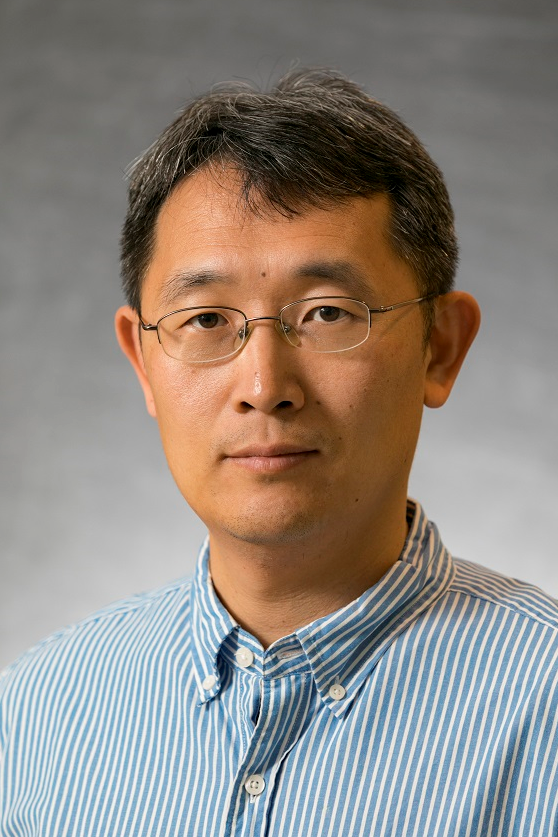}}]{Kai Sun}
(Fellow, IEEE) is a professor at the Department of EECS, University of Tennessee, Knoxville, USA. He received the B.S. degree in automation in 1999 and the Ph.D. degree in control science and engineering in 2004 both from Tsinghua University, Beijing, China. From 2007 to 2012, he was a project manager in grid operations and planning with the Electric Power Research Institute (EPRI), Palo Alto, CA. \end{IEEEbiography}

\begin{IEEEbiography}[{\includegraphics[width=1in,height=1.25in,clip,keepaspectratio]{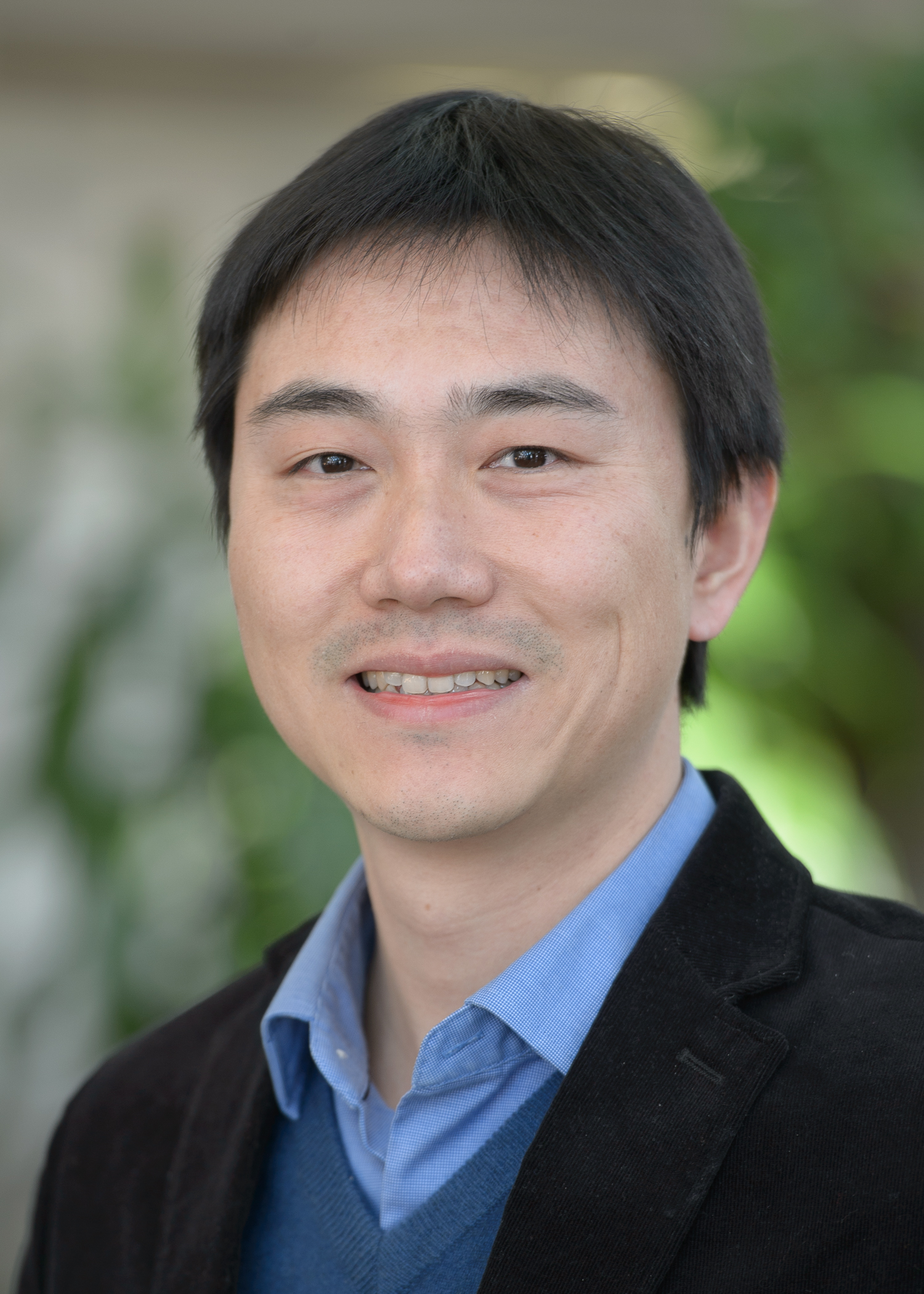}}]{Feng Qiu}
(Senior Member, IEEE) received his Ph.D. from the School of Industrial and Systems Engineering at the Georgia Institute of Technology in 2013. He is a principal computational scientist and a section leader with the Energy Systems Division at Argonne National Laboratory, Argonne, IL, USA. His current research interests include power system modeling and optimization, electricity markets, power grid resilience, machine learning and data analytics.\end{IEEEbiography}

\vfill

\end{document}